\documentclass{emulateapj}
\usepackage{natbib}
\usepackage{amssymb}
\usepackage{multirow}

\newcommand{\hone}{{{\sc H i}~}}

\slugcomment{To appear in the Astronomical Journal}

\shorttitle{\hone in the NGC2403 Group}
\shortauthors{Chynoweth et al.}

\begin{document}

\title{A Search for Diffuse Neutral Hydrogen and \hone Clouds in the NGC 2403 Group}

\author{Katie M. Chynoweth}
\affil{Vanderbilt University, Physics and Astronomy Department, 1807 Station B, 
Nashville, TN 37235}
\affil{National Radio Astronomy Observatory,
Green Bank, WV 24944}

\author{Glen I. Langston}
\affil{National Radio Astronomy Observatory,
Green Bank, WV 24944}

\author{Kelly Holley-Bockelmann}
\affil{Vanderbilt University, Physics and Astronomy Department, 1807 Station B, 
Nashville, TN 37235}

\and

\author{Felix J. Lockman}
\affil{National Radio Astronomy Observatory,
Green Bank, WV 24944}

\begin{abstract}
We have observed the NGC 2403 group of galaxies using the Robert C. Byrd Green 
Bank Telescope (GBT) in a search for faint, extended neutral hydrogen clouds 
similar to the clouds found around the M81/M82 group, which is located approximately 250 kpc from the
NGC 2403 group along the same filament of galaxies.
For an \hone cloud with a size $\leq 10$ kpc within 50 kpc of a group galaxy, 
our 7$\sigma$ mass 
detection limit is  2.2 $\times$ 10$^{6}$ M$_{\odot}$ for a cloud with a linewidth of 20 km
s$^{-1}$, over the velocity
range from -890 to 1750 km s$^{-1}$.
At this sensitivity level we detect 3 new \hone clouds in the direction of the group, as well as the known galaxies. 
The mean velocity of the new clouds 
differs
from that of
the group galaxies by more than 250 km s$^{-1}$, 
but are in the range of 
Milky Way High Velocity Clouds (HVCs) in that direction. It is most likely that the clouds are part of the Milky Way HVC population.
If \hone clouds exist in
the NGC 2403 group, their masses are less than 2.2 $\times$ 10$^{6}$
M$_{\odot}$.  We also compared our results to structures that are expected based on recent
cosmological models, and found none of the predicted clouds. If NGC 2403 is surrounded by a population of
dark matter halos similar to those proposed for the Milky Way in
recent models, our observations imply that their \hone content
is less than 1\% of their total mass.

\end{abstract}

\keywords{}

\section{Introduction}

\begin{figure}
\centering
\includegraphics[width=4in]{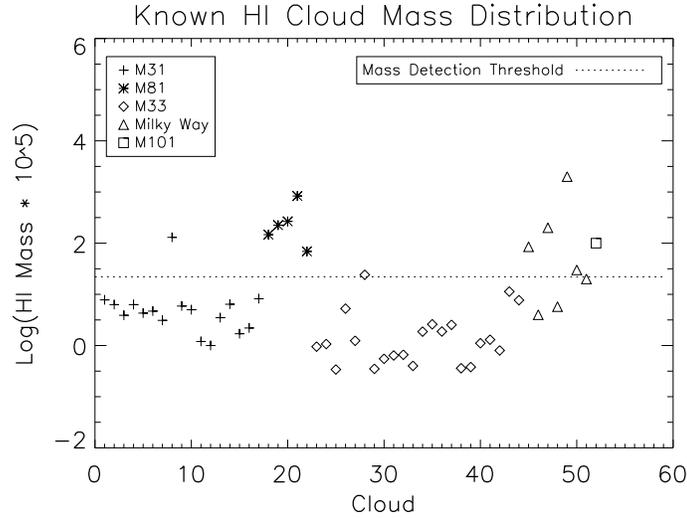}
\caption{Masses of known \hone clouds associated with nearby galaxies. Our mass
detection threshold is indicated with a dotted line. If this population of 
clouds is representative of the NGC 2403 group, our detection limit allows us to
detect the most massive 23\% of \hone clouds in the group. Data: \citet{thilker04},
\citet{grossi08}, \citet{chy08}, \citet{vdH88}, \citet{wakker07}, \citet{thom06},
\citet{simon06}, \citet{vanw99a}, \citet{put03}, \citet{loc08}, \citet{braun04}}
\label{massdist}
\end{figure}

The neutral hydrogen in the environment of galaxies and galaxy groups is 
morphologically and kinematically complex. Surrounding the Milky Way are the High Velocity Clouds
(HVCs), which are observed in both ionized and neutral states in large
complexes including the Magellanic Stream \citep{put03}, and as smaller isolated clouds, which together cover nearly 40\% of
the sky \citep{wvw97, loc02}. Possible analogues to the Milky Way clouds have been
 found around nearby galaxies including M31 \citep{thilker04}, M33 \citep{grossi08}, the M81/M82
 group \citep{chy08}, and M101 \citep{vdH88}. 
Figure \ref{massdist}
 shows for the mass distribution of extraplanar \hone clouds around nearby 
galaxies.  
 
There are many 
physical processess that could create  
HVCs and
extraplanar \hone clouds, most of which are likely to contribute to the overall population.
Some \hone clouds may be situated in dark halos -- clumps of dark matter and gas that have not yet formed
stars \citep{kra04}. \hone clouds may  
also be a product of cold-mode accretion along filaments (\citet{keres05}, Kere{\v s}
et al.~, private communication). Galaxy interactions can produce filamentary structures of 
gas and tidal streams like the Magellanic Stream and the tidal tails 
of the M81 group \citep{yun93, wal02}. Some \hone clouds may be the 
recently accreted by-products of galaxy formation \citep{mal04}, or may 
trace positions of faint dwarf galaxies \citep{bekhti06,
beg06}.  Some clouds may also arise from a ``galactic fountain", wherein hot 
gas created by supernovae expands thermally, cools, condenses and falls back 
into the galaxy \citep{bre80}.  Studies of
cloud properties such as mass, metallicity, morphology, and kinematics are 
necessary to distinguish between these possibilities within a given group. For example, to determine the relative importance of tidal interactions in generating \hone clouds, it is ideal to compare the \hone clouds within
nearby galaxy groups that are undergoing different levels of galaxy 
interactions.

\begin{figure}
\centering
\includegraphics[height=3in]{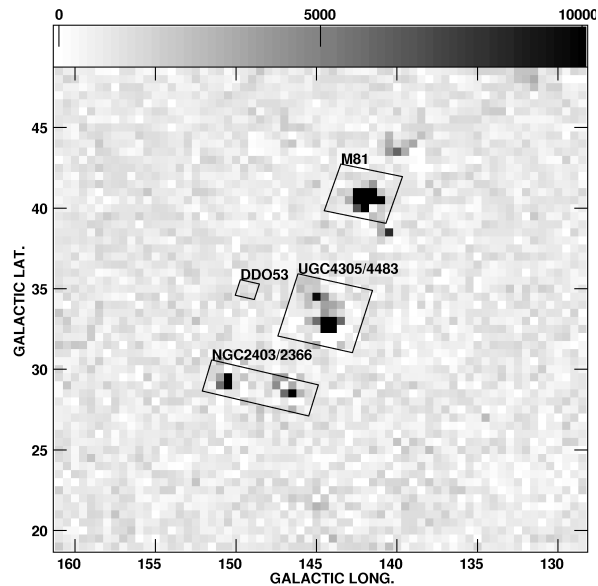}
\caption{Map of the NGC 2403 and M81/M82 groups along the M81 filament from the
Leiden/Argentine/Bonn Galactic HI survey \citep{kal05}, integrated from 107 to
200 
km s$^{-1}$. The grayscale range is 0 to 10000 K $\times$ m s$^{-1}$}
\label{filamentfig}
\end{figure}

GBT observations of the highly interacting M81/M82 group revealed a population of 
\hone clouds close to group galaxies, located along tidal streams, and
kinematically bound to the group \citep{chy08}.  
The \hone clouds in the M81 group are most likely byproducts of ongoing galaxy interactions. The NGC 2403 galaxy group is located along the same filament of galaxies as the
M81 group -- the ``M81 filament" \citep{kar02}, shown in Figure
\ref{filamentfig}. The NGC 2403 group contains the bright 
spiral galaxy NGC 2403, 3 additional large galaxies, and at least 4 dwarf or 
low surface brightness satellite galaxies. Table 1 lists the properties of the 
group. In contrast to the M81/M82 group, the NGC 2403 group is not known to be
interacting at all. Although the NGC 2403 group appears mostly quiescent and
undisturbed, some of the group galaxies show subtle irregularities which
may be due to previous interactions between group galaxies.  
The
largest galaxy, NGC 2403, shows a thick, lagging \hone layer and a few \hone filaments with anomalous velocities,
which are similar to the M81/M82 clouds in mass \citep{frat02}. These features ma be
analogues to Milky Way HVCs, and their origins are similarly unclear.  
These features have been attributed to a combination of outflow 
from a galactic fountain and accretion from the IGM \citep{frat06}; both of 
these phenomena may be induced by interactions between group galaxies
\citep{lar78, san08}.  UGC 4305 has a cometary appearance in \hone \citep{bureau02}, which may be caused 
by ram pressure from the intragroup medium. The dwarf galaxy DDO 53 is also kinematically irregular in
\hone \citep{beg06}. The other galaxies in the group appear to be normal. 
Galaxy distances and velocities are taken from \citet{kar07}.

We observed the NGC 2403 group with
the GBT to map its faint, extended neutral gas and to generate a 
census of its neutral hydrogen cloud population. 
A survey of the \hone cloud population in this quiescent group, combined 
with the results from the M81/M82 group, allows us to explore the role 
that interactions play in the formation and evolution of 
extraplanar \hone clouds.  

In sections \ref{obsec} and \ref{datasec} we summarize the observations and data reduction. In section \ref{interp} we describe methods used to interpret our observations. In 
section \ref{knownmsec}, mass calculations for known objects in the group are presented. In section
\ref{ressec} we discuss the properties of the newly detected structures.  Finally, in section \ref{consec} we discuss the implications of our results 
in light of recent models of the formation of the Milky Way.

\section{Observations}
\label{obsec}
We observed the NGC 2403 group with the 100m Robert C. Byrd Green Bank Telescope (GBT) of the 
NRAO\footnote{The National Radio Astronomy Observatory (NRAO) is a facility of the 
National Science Foundation operated under cooperative agreement by Associated Universities, Inc.}
in 9 sessions between 8 June and 1 August, 2008.  We combined the 9 observing
sessions into three maps: a $6^{\circ} \times 2^{\circ}$ 
region (``Region 1") containing NGC 2403, NGC 2366, and DDO 44; a $4^{\circ}
\times 4^{\circ}$ region (``Region 2")
containing UGC 4305, UGC 4483, and K52, with an extension to observe the region
around VKN; and a $1^{\circ} \times 1^{\circ}$ region around DDO 53. Figure
\ref{filamentfig} shows a layout of the
observations. Additionally, in light of the detection of a cloud at 07$^h$36$^m$12.1$^s$, 67$^{\circ}$33$^{``}$25$^{`}$, we 
re-observed the $1^{\circ} \times 1^{\circ}$ region centered on the object for
an additional 22 minutes.  

The regions were observed by moving the telescope in declination and sampling every $3\arcmin$ at an integration time of 3 seconds per sample.  Strips of constant declination were 
spaced by  $3\arcmin$.  Corresponding maps were made by moving the telescope in right ascension to form a `basket weave' pattern over the region. 
In the end, each $3\arcmin$ pixel was integrated for approximately 6 seconds 
except in the region surrounding the large new cloud, where each pixel 
received 18 seconds.

The GBT spectrometer was used with a bandwith of 12.5 MHz, corresponding to a
velocity range of -890 to 1750 km s$^{-1}$. The typical system temperature for each channel of the dual-polarization 
receiver was $\sim$ 20 K.

Our observations cover a large angular area (at least 50 kpc in projection around each group
galaxy) in order to make a complete study of 
the \hone cloud population and properties across the entire extent of the group, 
including between major galaxies. The beam size of the GBT in the 21 cm line is
$9\arcmin$, which maps to approximately 10 kpc at the distance of the NGC 2403
group; this is well 
matched to the expected angular size of \hone clouds above our mass detection
threshold \citep{thilker04, chy08}.

\section{Data Reduction}
\label{datasec}

The GBT data were reduced in the standard manner using the GBTIDL and AIPS\footnote{Developed by the National Radio Astronomy Observatory;
documentation at http://gbtidl.sourceforge.net, http://www.aoc.nrao.edu/aips} 
data reduction packages.

In order to match our velocity resolution to the expected linewidths of \hone 
clouds in the group, spectra were smoothed  
to a channel spacing of 24.4 kHz, corresponding to a velocity resolution of 
5.2 km s$^{-1}$. A reference spectrum for each of the 9 observation sessions 
was made using an 
observation of an emission-free region, usually from the edges of the maps. The reference spectrum was then used to 
perform a (signal-reference)/reference calibration of each pixel. 
The calibrated spectra were scaled by the system temperature, corrected 
for atmospheric opacity and GBT efficiency. We adopted GBT efficiency equation 
(1) from \citet{lang07} with a zenith atmospheric opacity 
$\tau_{0}$ = 0.009. Velocities are in the kinematical LSR 
velocity frame. 

The frequency range observed was relatively free of RFI and less than 0.3 \% of 
all spectra were adversely affected. The spectra exhibiting RFI values were 
identified by tabulating the RMS noise level in channels free of neutral 
hydrogen emission.  Spectra that showed high values of RMS
noise across many channels were flagged and removed.  
Observations were gridded using the AIPS task SDIMG, which also averages 
polarizations. After amplitude calibration and gridding, a 3rd-order polynomial 
was fit to line-free regions of the spectra. This baseline was subtracted from 
the gridded spectra. 
The effective angular resolution, determined 
from maps of 3C286, is 9.15\arcmin $\pm$ 0.05\arcmin. 
To convert to units of flux density, we observed the calibration source 3C286, 
whose flux density is 14.57 $\pm$ 0.94 Jy at 1.418 GHz  
\citep{ott94}. The calibration from 
K to Jy was derived by mapping 3C286 in the same way that the
 \hone maps were produced. 
After all corrections for the GBT efficiency and the mapping process, the 
scale factor from K/Beam to Jy/Beam images is 0.43 $\pm$ 0.03. 
The typical RMS noise in the final data cube is 
 25 mK per 24.4 kHz channel.  The instrumental parameters are 
summarized in Table 2, and Table 3 gives a summary of the observations.

\section{Data Interpretation}
\label{interp}

The mapped regions have projected sizes of 122 $\times$ 367 kpc$^{2}$ for Region
1, 245 $\times$ 245 kpc$^{2}$ for Region 2, and 61 $\times$ 61 kpc$^{2}$ for DDO
53. Based on previous studies, which found extraplanar clouds  within 1-50 kpc from the center 
of a galaxy \citep{wvw97,thilker04,chy08}, we should detect all of the \hone clouds around group galaxies that
occupy the same phase space as the clouds around M31, the M81 group, and the Milky Way.

Previous studies have found clouds only within 150 km s$^{-1}$ of the 
systemic velocity of associated galaxies \citep{thilker04, chy08}.  If these values can be interpreted as
a prediction for \hone cloud locations, we expect to find clouds with velocities
within a few hundred km s$^{-1}$ 
relative to the main galaxies in the group. Spectral maps were made in the entire observed velocity range, from -890 to 1750 km s$^{-1}$.  This extends the search more
than $\pm$ 1000 km s$^{-1}$ beyond the systemic velocity of any group galaxy. Spectral maps were visually inspected for extended emission not associated with group
galaxies and possible new \hone clouds. The mass detection threshold was
calculated assuming that clouds would be unresolved in the GBT beam, using the
formula
\begin{equation}
\left(\frac{\sigma_M}{M_{\odot}}\right)=2.36 \times 10^5 \left(\frac{D}{Mpc}\right)^2 \left(\frac{\sigma_s}{Jy}\right)
\left(\frac{\Delta V}{km\hskip 1mm s^{-1}}\right) \sqrt N
\end{equation}

Where $D$ is the distance to the group, $\sigma_s$ is the rms noise in one channel, $\Delta V$
is the channel width, and $N$ is the number of channels required for a secure detection. For our calculation, we used a \hone cloud linewidth of 
20 km s$^{-1}$, corresponding to $N=4$.

The 7$\sigma$ detection threshold for \hone mass was  
2.2 $\times$ 10$^{6}$ M$_{\odot}$, adopting the NGC 2403 distance of 3.3 Mpc.
With this mass detection threshold, we would be able to detect analogues to
the most massive clouds around M31 and M33, the M81/M82 clouds, the M101 cloud, 
and large Milky Way objects such as Complexes C and H and the Magellanic Stream.
These clouds are the most likely to originate in galaxy interactions
and/or primordial accretion, rather than from supernovae. Figure \ref{massdist}
shows that if the mass spectrum of extraplanar \hone clouds associated with the Milky Way and nearby
galaxies is representative, we should detect 23\% of the extraplanar \hone cloud population in the 
NGC 2403 group. 

The spectral line images are available on-line\footnote{The spectral line image
cubes are available http://www.gb.nrao.edu/$\sim$kchynowe/N2403/}.

\section{Masses of Known Objects}
\label{knownmsec}

\begin{figure}
\centering
\begin{tabular}{cc}
\includegraphics[width=2.5in]{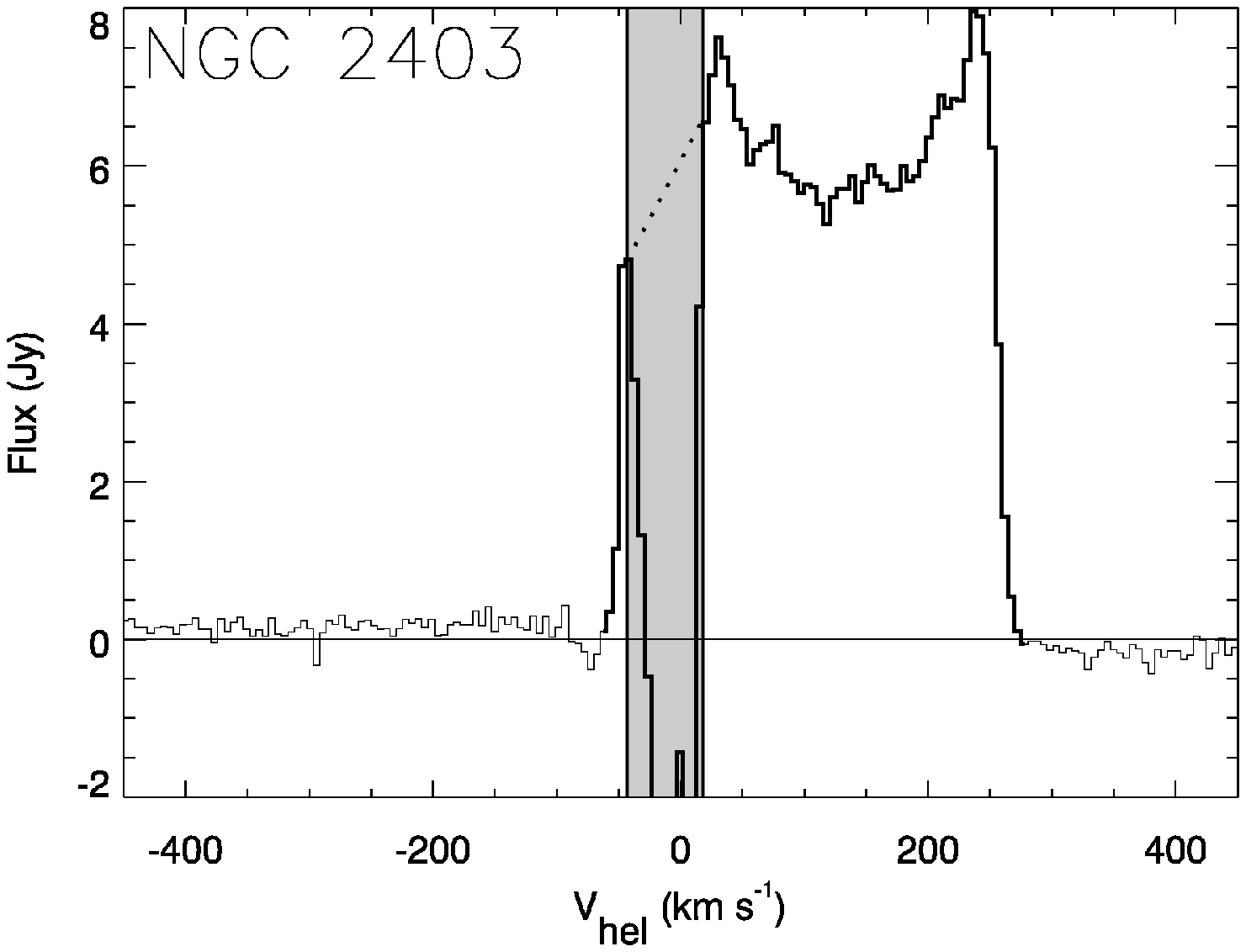}  &
\includegraphics[width=2.5in]{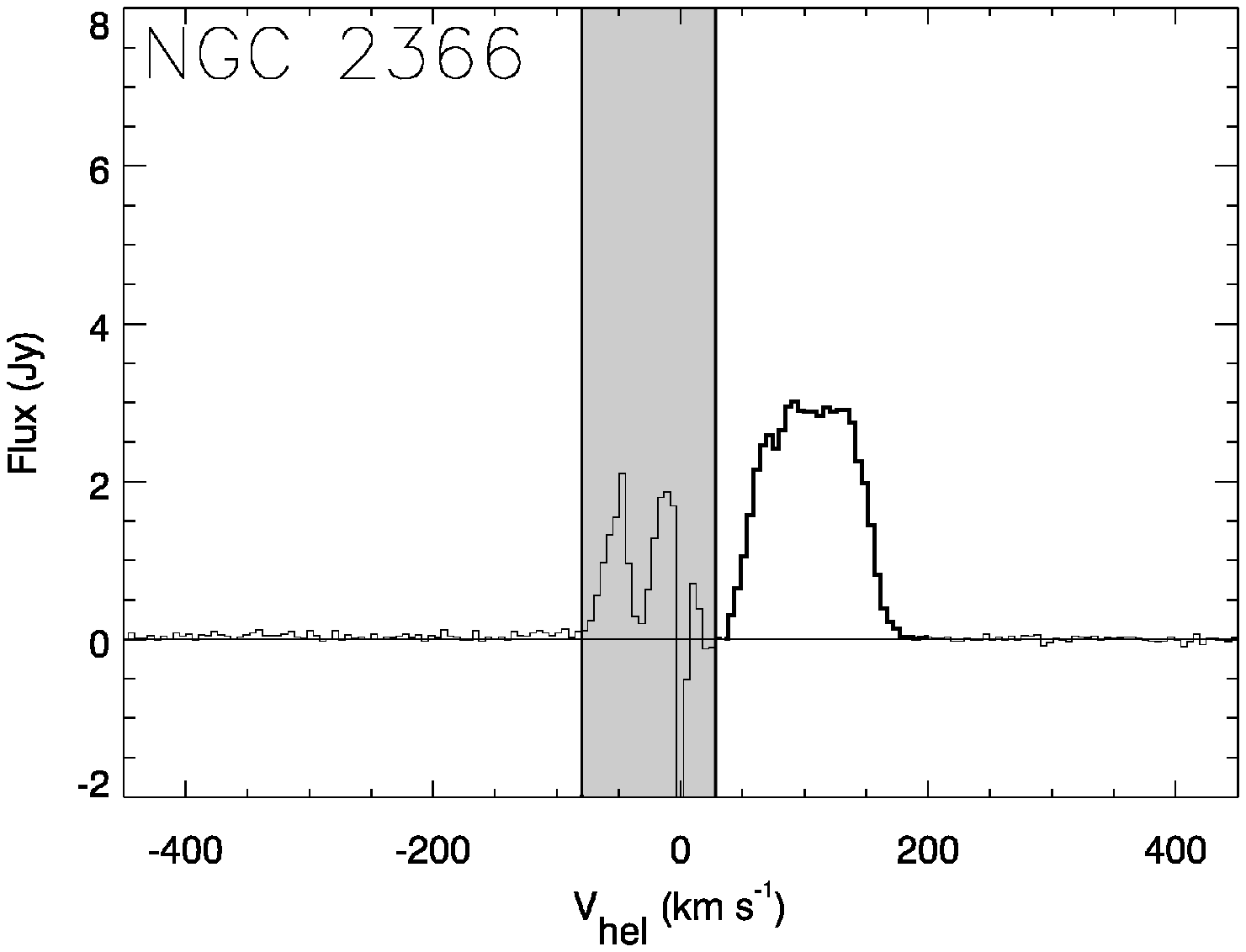}  \\
\includegraphics[width=2.5in]{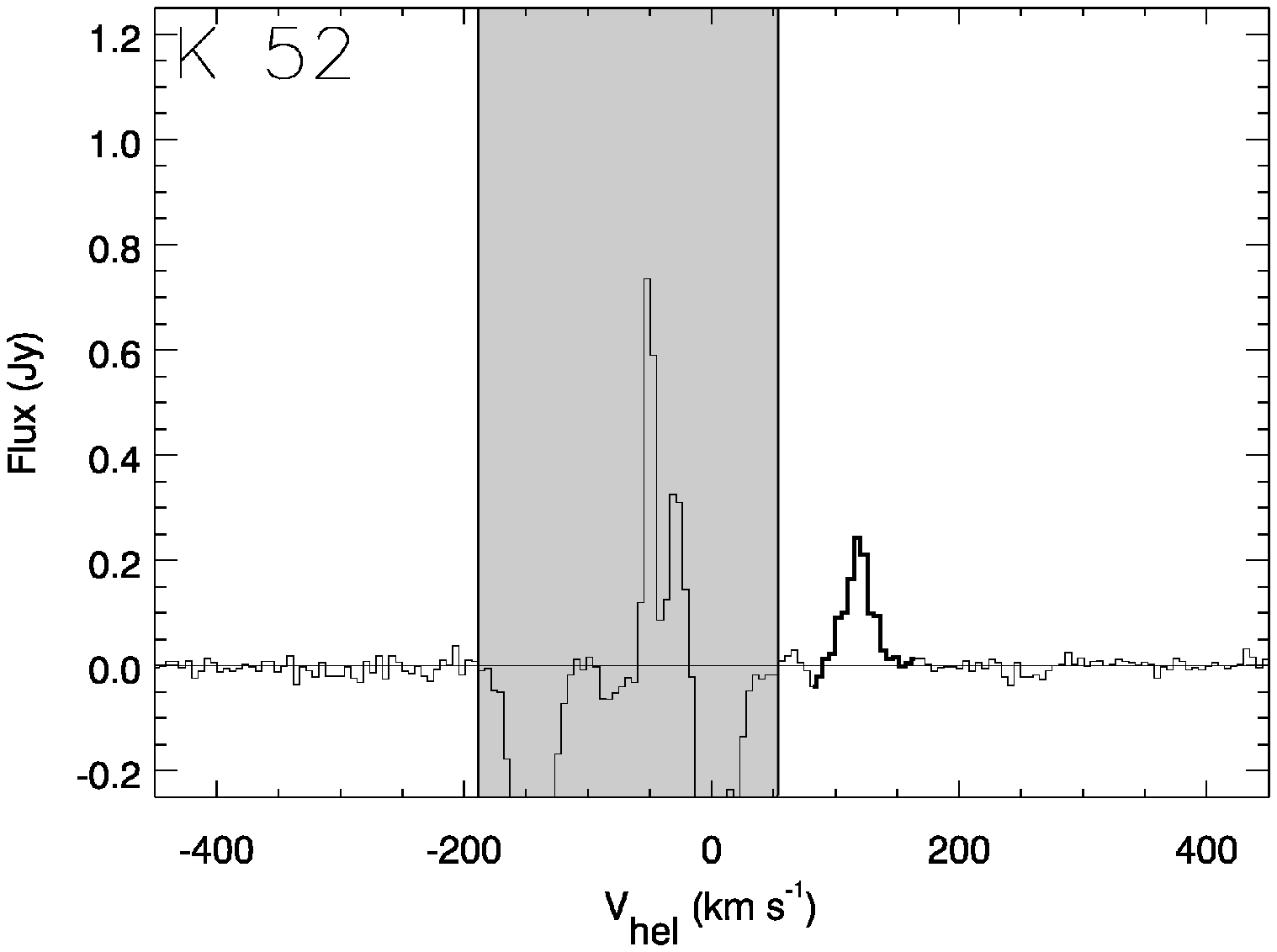}  &
\includegraphics[width=2.5in]{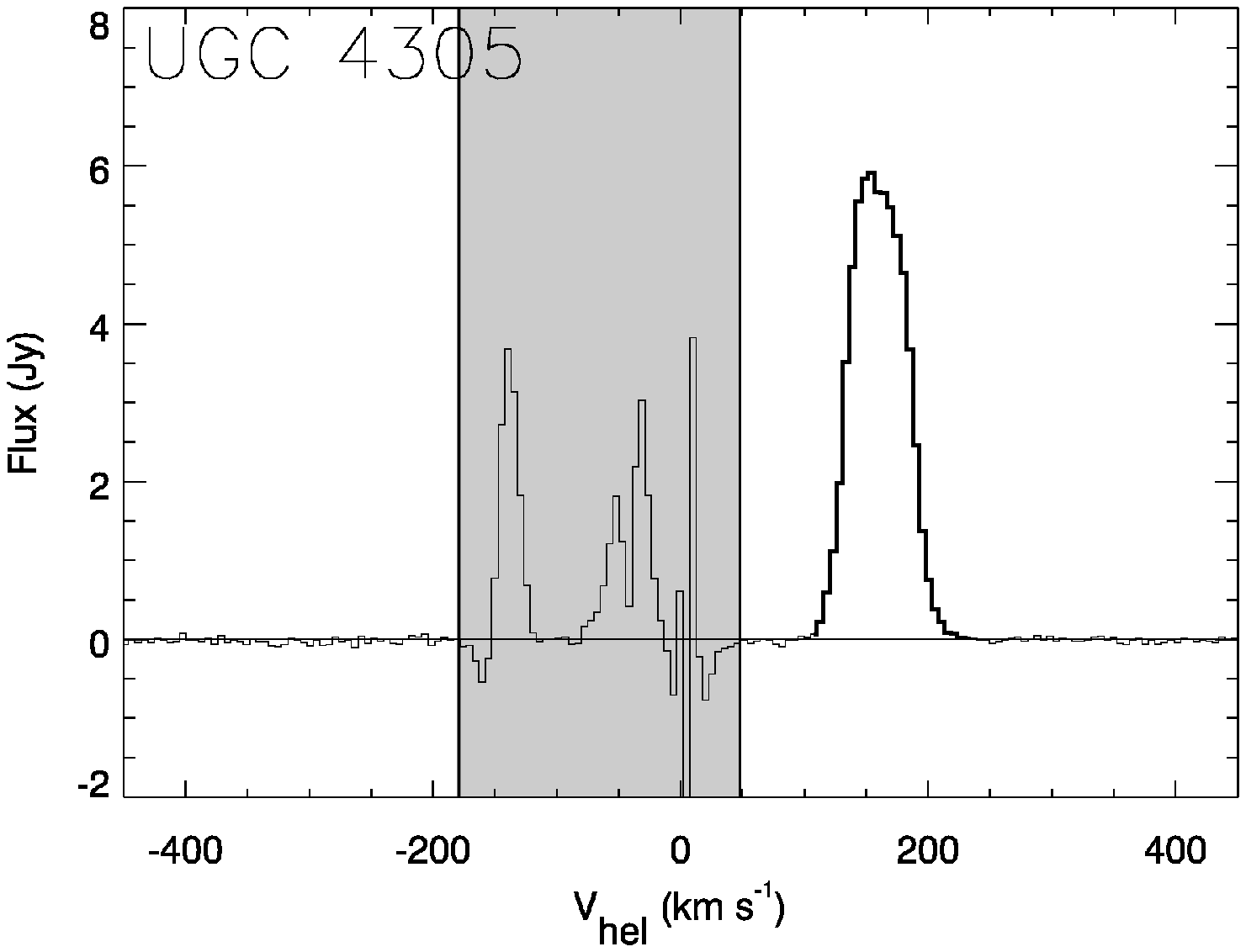}  \\
\includegraphics[width=2.5in]{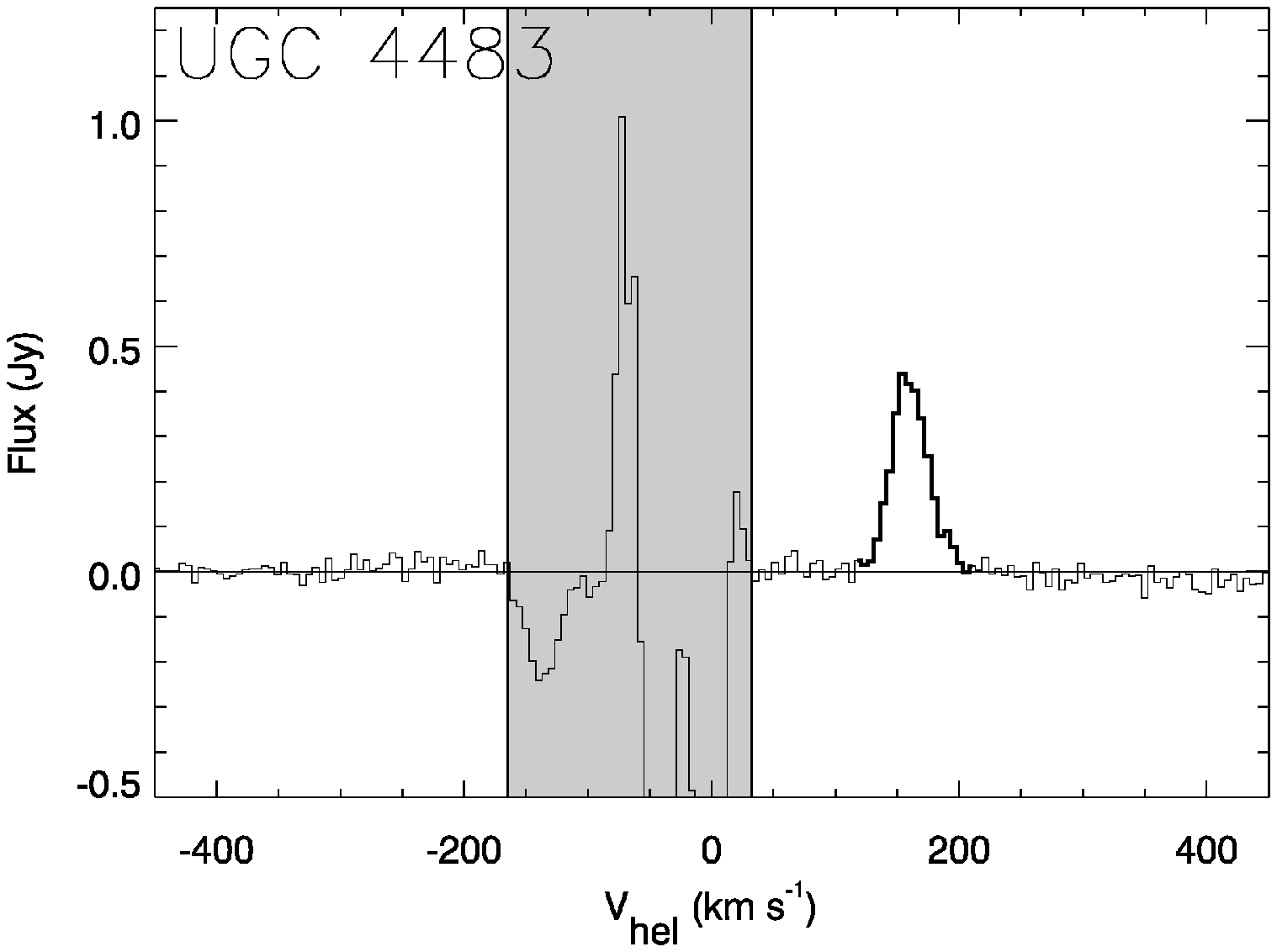} & 
\includegraphics[width=2.5in]{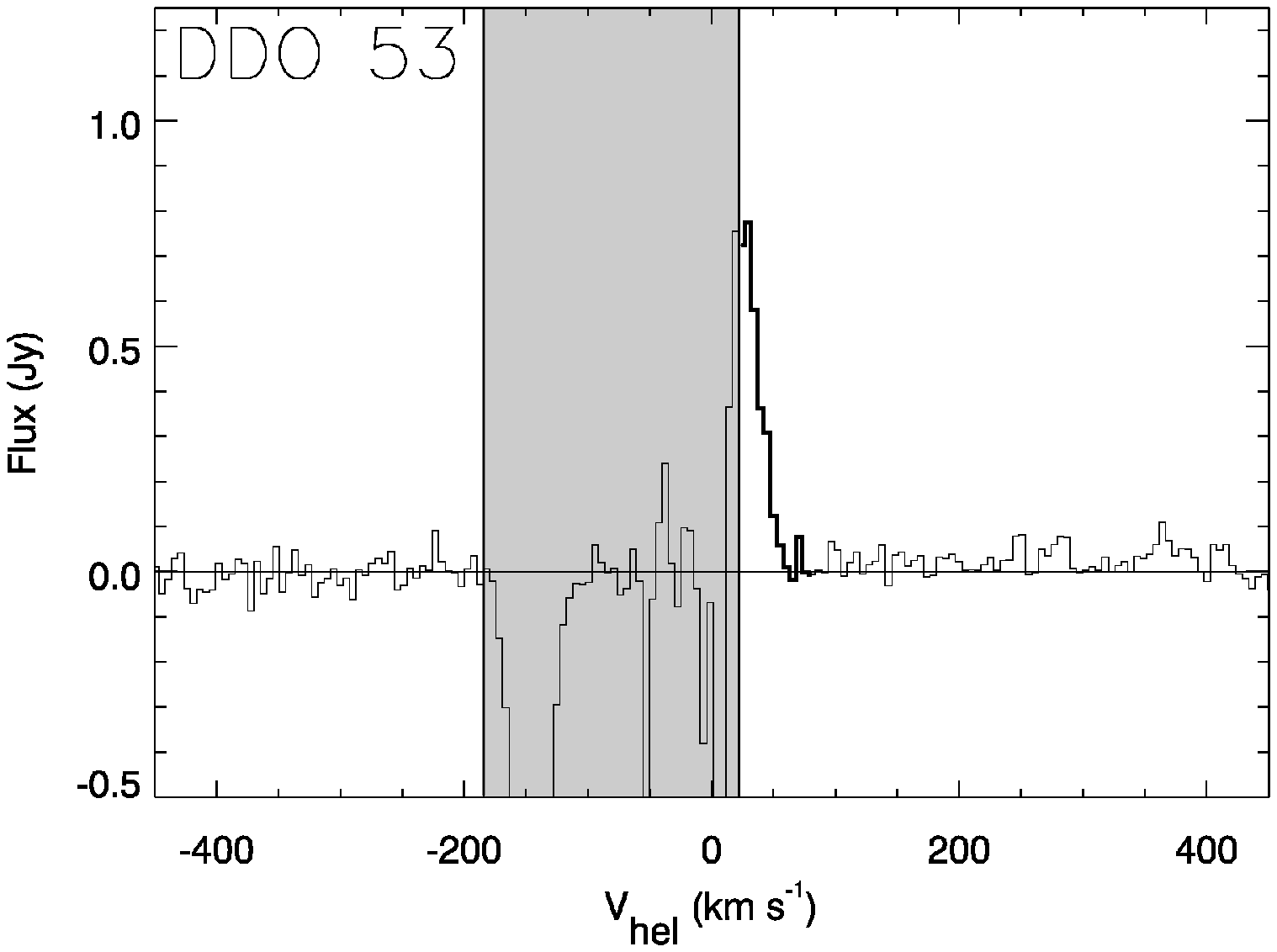} \\
\end{tabular}
\caption{\hone intensity profiles of major group galaxies. The regions of the
spectrum used to calculate \hone mass are highlighted in bold. From top left:
NGC 2403, NGC 2366, K 52, UGC 4305, UGC 4483, and DDO 53. The velocity
ranges contaminated by foreground gas are indicated with gray boxes.}
\label{glx_ispec}
\end{figure}

 Moment maps of the \hone  
column density are presented in Figures \ref{mom01} and \ref{mom02} and 
the \hone spectra for the known galaxies are presented in Figure
\ref{glx_ispec}. The velocity ranges affected by foreground Milky Way emission
are indicated on each plot. 
There is some overlap in velocity space between the NGC 2403 group and the 
Milky Way, and over this velocity range we can not discriminate between local 
and distant emission. Of the galaxies, only NGC 2403 overlaps with Milky Way
velocities. To estimate the 
total \hone mass of neutral hydrogen for NGC 2403, we replaced the velocity channels contaminated by Milky Way 
and local high velocity clouds with interpolated values computed from spectral 
line observations at either end of the contaminated channels.

Our calculated 
values for $M_{HI}$ are in agreement with the literature; see Table 4. Error estimates on the masses are 1$\sigma$. The dominant error 
contribution is the uncertainty in the absolute calibration, about 7\% for the
GBT.
 
\section{New Identifications} 
\label{ressec}

All of the galaxies in the NGC 2403 group appear 
undisturbed at the angular resolution of the GBT. We do not detect the filaments of \hone in NGC 2403
seen in higher-resolution 
VLA data \citep{frat02}, most likely due to beam dilution.

\begin{figure}
\centering
\begin{tabular}{cc}
\includegraphics[width=2.5in]{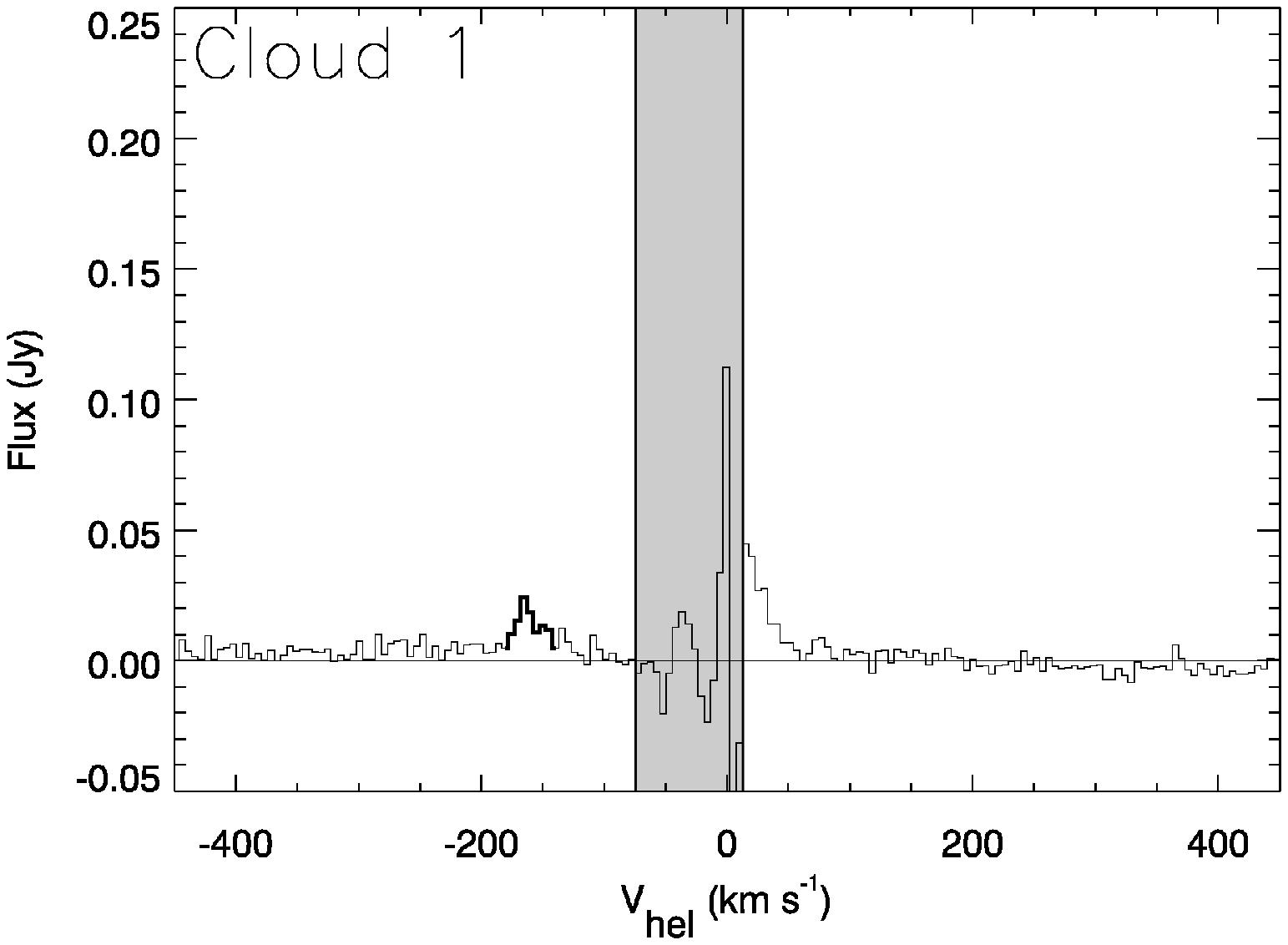} &
\includegraphics[width=2.5in]{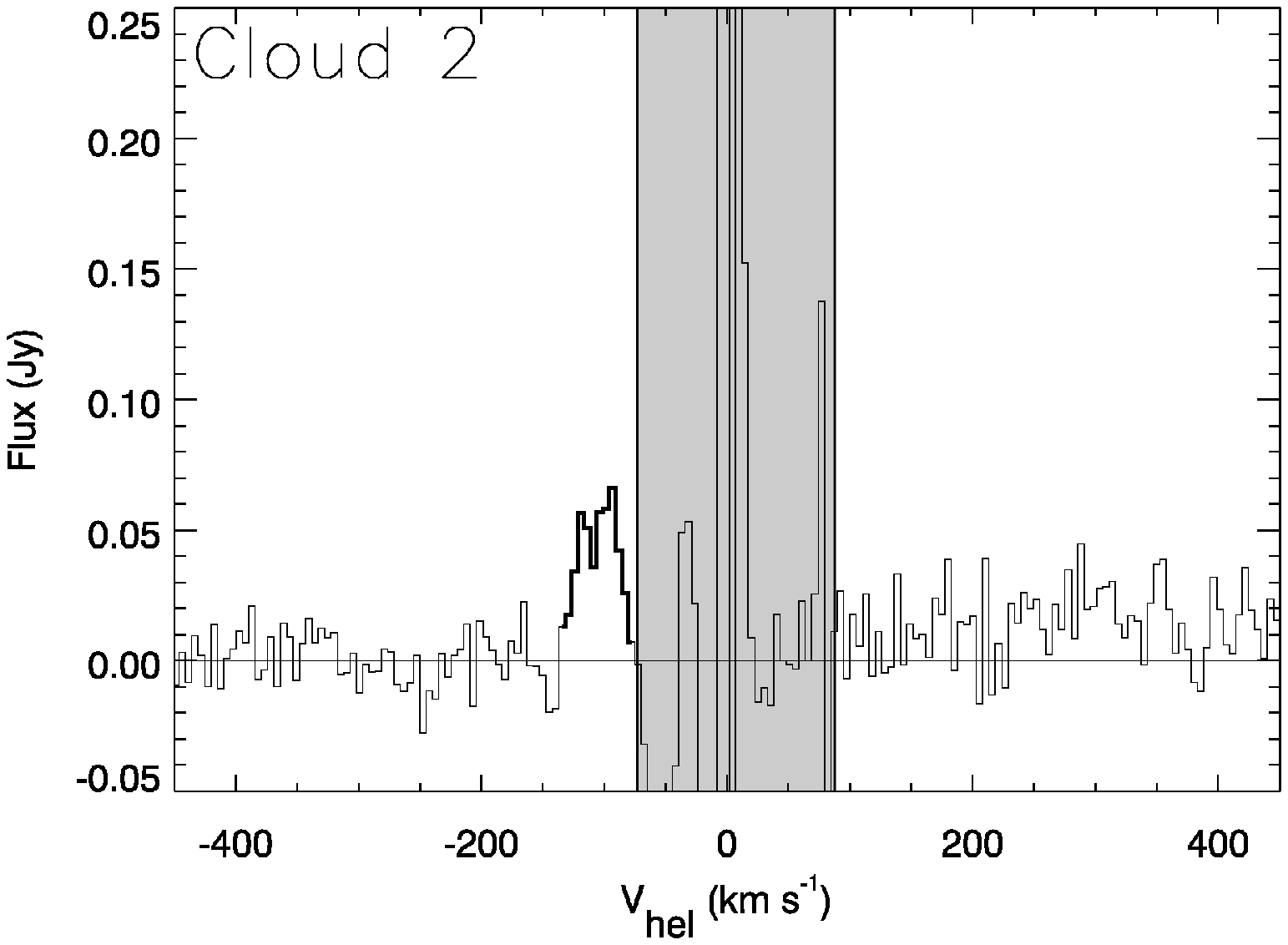} \\
\includegraphics[width=2.5in]{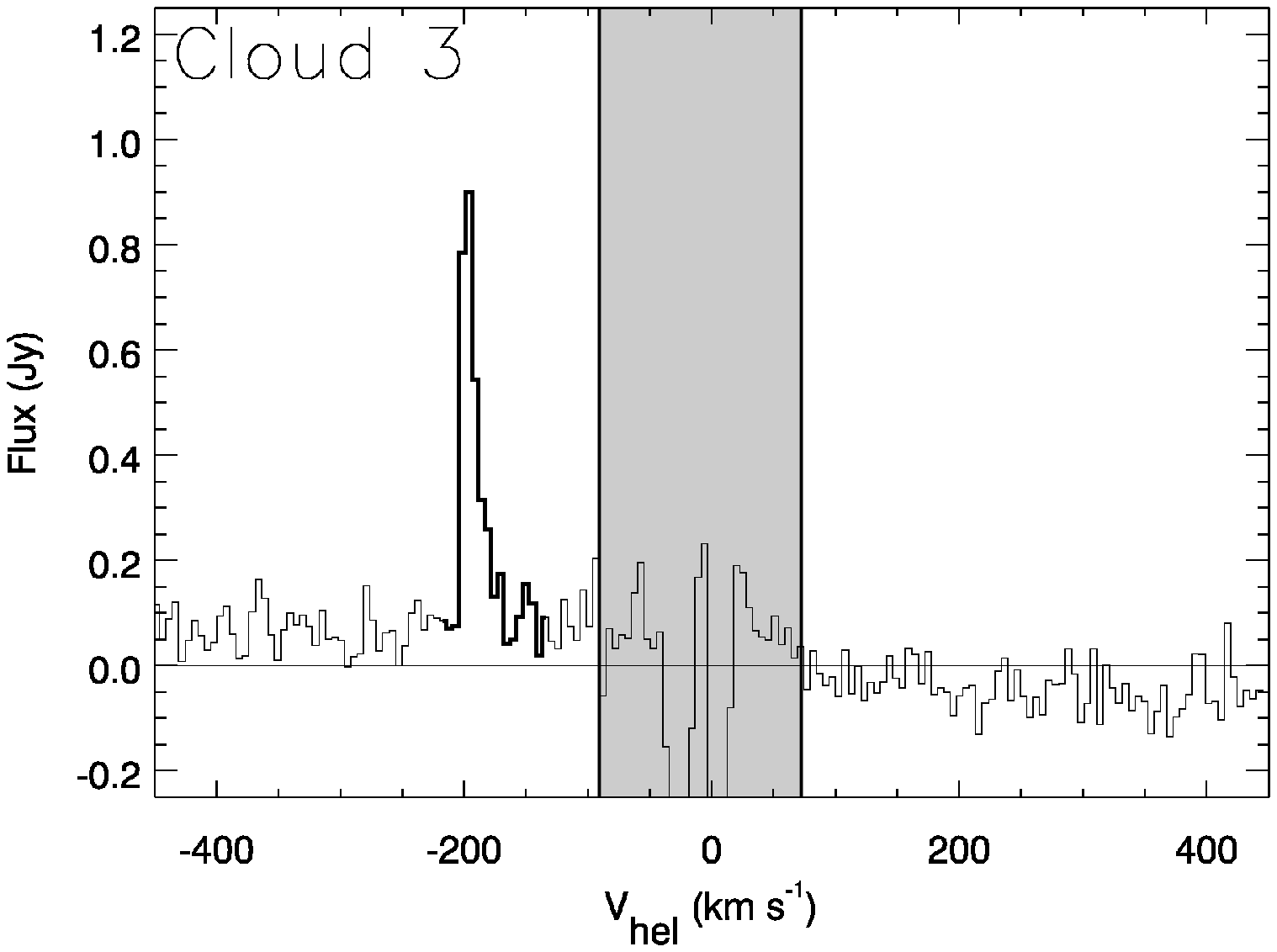} & \\
\end{tabular}
\caption{\hone intensity profiles of Clouds 1, 2, and 3. The regions of the
spectrum used to calculate \hone mass are highlighted in bold.  The velocity
ranges contaminated by foreground gas are indicated with gray boxes.} 
\label{cloudspec}
\end{figure}

\begin{figure}
\centering
\includegraphics[width=3.5in]{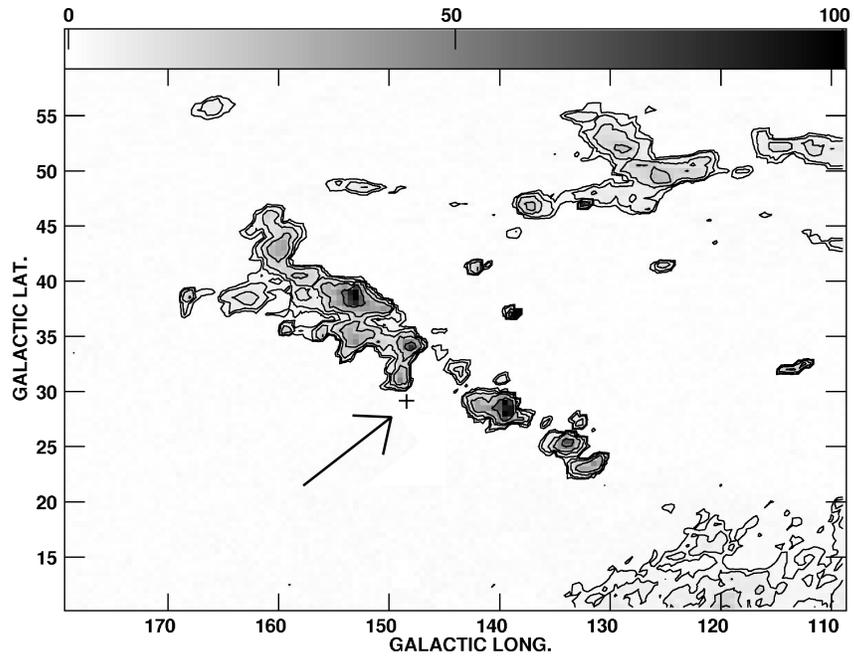}
\caption{\hone total intensity map of HVC Complex A from the LAB survey of
Galactic \hone \citep{kal05}.
Integrated from -250 to -150 km s$^{-1}$. The location of the NGC 2403 group GBT observations is marked with a cross at approximately
l=150$^{\circ}$, b=30$^{\circ}$.}   
\label{laba}
\end{figure}

Three \hone clouds are detected near NGC 2403 and NGC 2366, labeled numbers 1-3 in Figures 
\ref{mom01} and \ref{mom02}. The clouds are described below, and Table 5 lists
their properties. Spectra of the clouds are shown in Figure \ref{cloudspec}. 
It is not
immediately clear whether the clouds are extragalactic or local to 
the Milky Way. The region mapped is
just south of the Milky Way HVC Complex A (see Figure \ref{laba}), so the clouds
may be associated with Complex A or other Milky Way HVCs. In order to
distinguish between the two possibilities, we analyzed the spectra of the clouds.

Each cloud is 
detected in 4-6 channels, corresponding to velocity widths of 20-30 
km s$^{-1}$ and consistent with both a local and an extragalactic origin \citep{bekhti06,west08}. 
If the clouds are 
located at the distance of NGC 2403 (3.3 Mpc), their masses are 1.5 to 51.7
 $\times$ 10$^6$ M$_{\odot}$ and their diameters range from $\sim$ 9-25 kpc. If, instead, 
the clouds are within the distance bracket for Complex A given in \citet{vanw99a}, 
their masses are between 2-475 M$_{\odot}$ and their diameters are between 10-75 pc.
Either of these sets of numbers is plausible: the \hone clouds discovered in the
M81/M82 group had masses $\sim$ 10$^7$ M$_{\odot}$ and diameters $\sim$ 10 kpc; likewise,
Milky Way \hone clouds are found with masses  $\sim$ 60
M$_{\odot}$ and sizes $\sim$ 10 pc \citep{stil06}. 

\begin{figure}
\centering
\begin{tabular}{cc}
\includegraphics[width=2.5in]{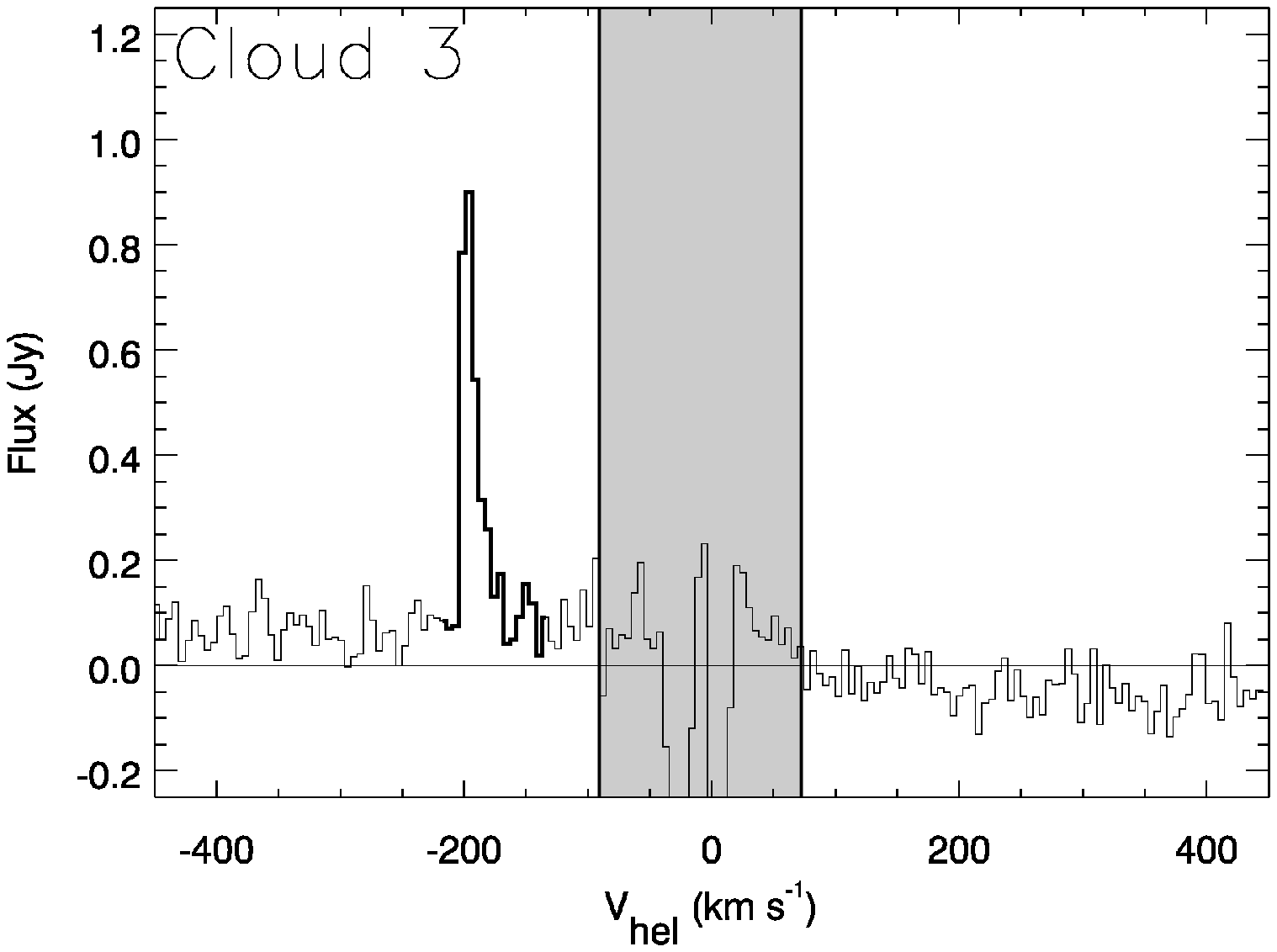} &
\includegraphics[width=2.5in]{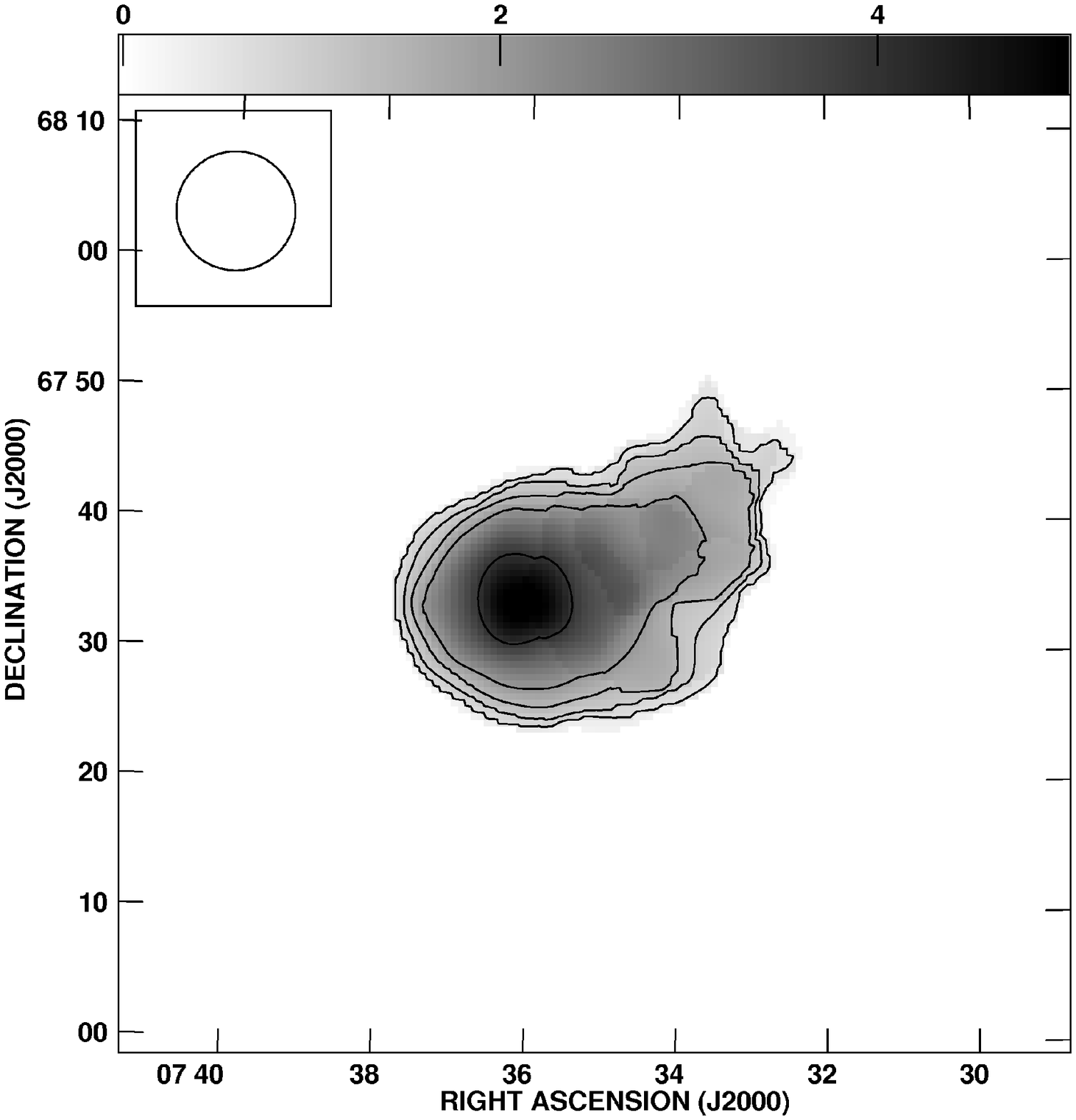} \\
\includegraphics[width=2.5in]{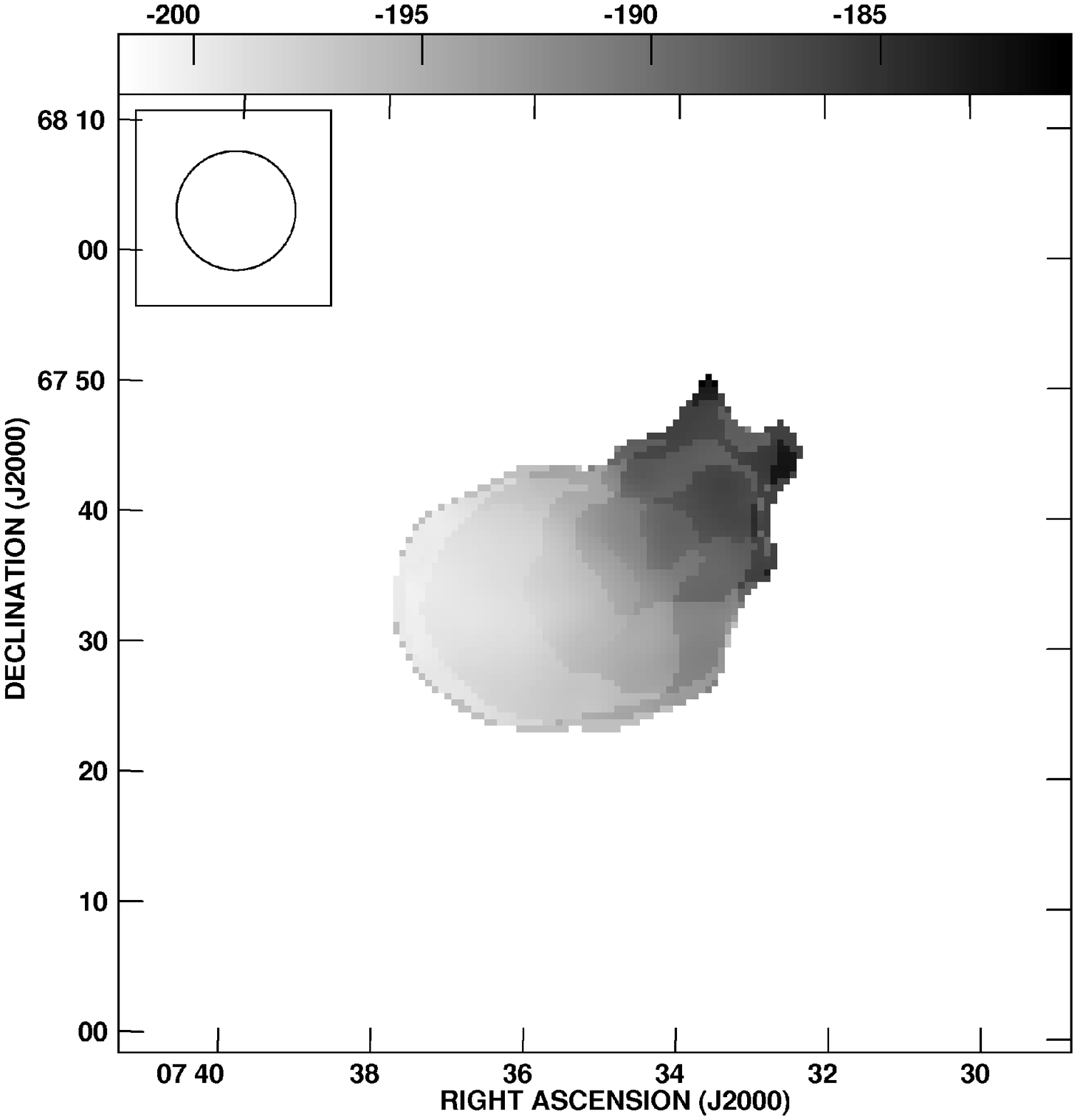} &
\includegraphics[width=2.5in]{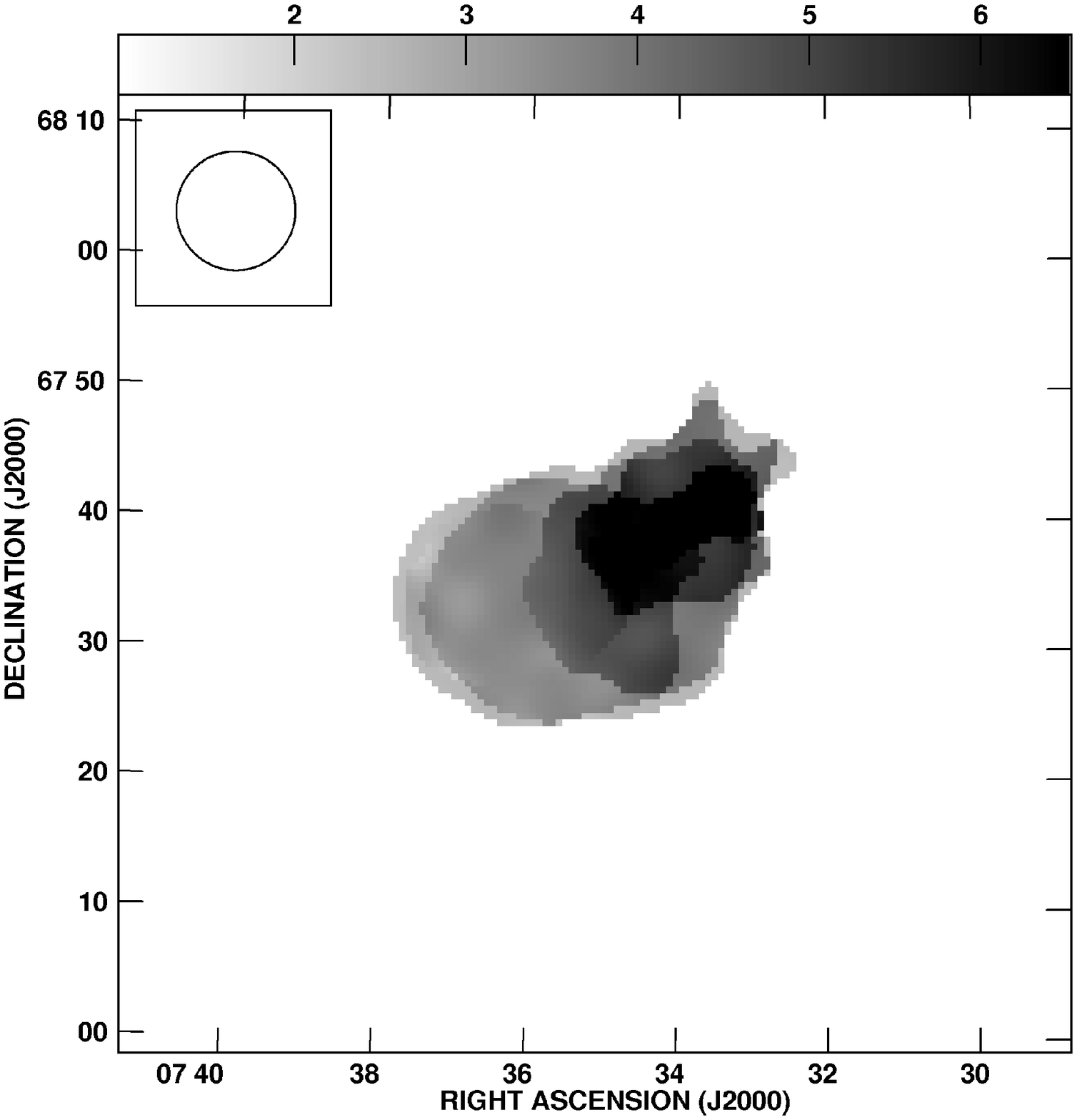} \\
\end{tabular}
\caption{The compact clump of gas  ``Cloud 3". From top left: \hone intensity 
profile, \hone column density (grayscale: 0 to 5 Jy beam$^{-1}$ $\times$ km s$^{-1}$), velocity
(grayscale: -200 to -180 km s$^{-1}$), velocity dispersion (grayscale: 1 to 6.5 km s$^{-1}$). Maps are integrated from -240 to
-160 km s$^{-1}$.}
\label{new}
\end{figure}

Clouds 1 and 2 are unresolved in the GBT beam. However, the structure of Cloud 3 was
further analyzed. The cloud has the
cometary or `head-tail' structure, velocity gradient, and velocity-dispersion gradient 
observed in many Milky Way HVCs \citep{bekhti06}. These features are all
presented in Figure \ref{new}. This type of structure is indicative of ram pressure interaction between a moving cloud and an ambient medium, i.e.\ the
warm gaseous halo of a galaxy \citep{qui01}.
If at the distance of the NGC 2403 group, Cloud 3 is located $\ge$ 90 kpc from
any group galaxy, and is far from the outer regions of these galaxies. 
It is thus likely that the cloud is part of
the Milky Way, and interaction with the Milky Way halo is the source of 
its head-tail structure.

\begin{figure}
\centering
\includegraphics[width=4in]{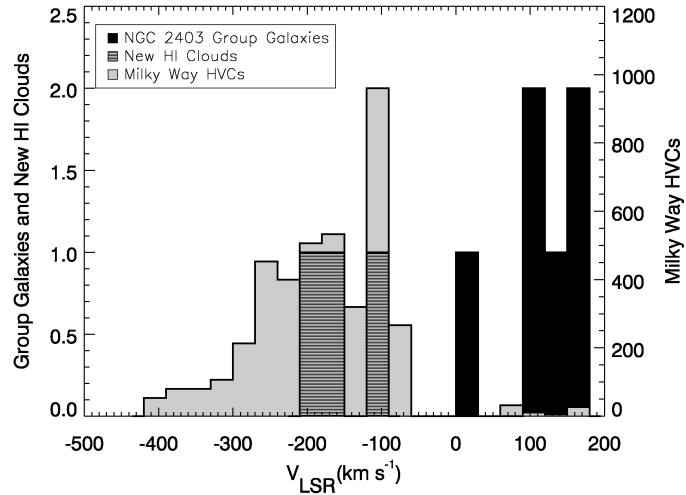}
\caption{Velocity distribution of NGC 2403 group galaxies, new \hone cloud detections, and Milky Way HVCs in the direction of the NGC
2403 group from \citet{vanW04}. The new \hone clouds have velocities consistent with the Milky Way HVCs, but inconsistent with the group galaxies.}
\label{vhist}
\end{figure}

We also compared the velocity distributions of the new \hone clouds, the
group galaxies, and the Milky Way HVCs in the direction of the group; see
 Figure \ref{vhist}. If
the new clouds are part of the NGC 2403 galaxy group, their
velocity distribution ought to be consistent with that of the galaxy group
\citep{chy08}. The average cloud velocity is -142 km s$^{-1}$, 
separated by 250 km s$^{-1}$ from the NGC 2403 galaxy group average
velocity of 113 km s$^{-1}$. The velocity
distribution of the clouds is consistent with the velocity of Milky Way HVCs in that direction
\citep{vanW04}.   Therefore the
clouds are likely to be newly discovered compact components of Complex A or other compact
Milky Way HVCs rather than members of the NGC 2403 group.

\section{Discussion and Conclusions}
\label{consec}

We have observed the NGC 2403 galaxy group with the GBT in order to search for
faint, extended neutral hydrogen and \hone clouds. We report 
detection of three clouds above our 7$\sigma$ detection 
threshold of 2.2 $\times$ 10$^{6}$ M$_{\odot}$ for
sources filling the 9$\arcmin$ GBT beam. These clouds are separated in velocity from the galaxy group by 250 km s$^{-1}$. The clouds are
consistent in position with Milky Way
HVC Complex A and consistent in velocity with all
Milky Way HVCs in the direction of the group.
We conclude that the clouds are most likely 
associated with the Milky Way, and that we 
have detected no extraplanar \hone clouds associated with the NGC
2403 galaxy group.
Any \hone clouds in the group must be less massive than 2.2 $\times$ 10$^{6}$ M$_{\odot}$, or lie farther than 50 kpc from a group galaxy.

Cosmological galaxy formation models predict a population of dark halos; clumps of dark matter and gas that have not yet formed stars.   
These dark halos have not been observed, but 
\citet{kra04} point out that some of the missing
dark halos may be the population of compact HVCs identified by 
\citet{braun99}.
The existence of any HVCs associated with dark halos remains in doubt
and our observations of nearby galaxy groups are designed to
place strong limits on the existence of this class of object.
We did not find any clouds in the NGC 2403 galaxy group with the
properties expected for the missing dark halos.  

These observations are significant because of the small number of
clouds (none) detected with the properties expected for galaxy
formation based on cold dark matter models.  If \hone clouds
indeed form within dark halos, it would be useful to determine
how many of these halos our observations should have been able to detect.
Recently, \citet{die08} performed a detailed numerical model
of the Milky Way formation based on cosmology, called Via Lactea II
(see http://www.ucolick.org/$\sim$diemand/vl/ for more details).
The public release of this simulation consists of a table of
resolved sub-halos, with the total masses and 3 dimensional positions
and velocities.  In order to connect their simulations with our
observations we require an estimate of the ratio of \hone mass to dark
matter.  \citet{mal03}
have modeled the ratio of gas to dark matter at various densities
and under the influence of ionizing photon flux.   Their models
suggest a range of neutral \hone to dark matter ratios,
from 0.015 to 0.0015, depending on the gas density.
We assume a \hone to dark matter ratio of 0.01, and
compute the number of clouds that could be detected by our observations.
We have taken the Via Lactea II simulation and applied our
observing selection criteria of mass limits, angular region surveyed
and velocity ranges searched, and computed the number of dark matter
halos we would detect, if each halo has 1\% of its mass in \hone.

\begin{figure}[ht]
\begin{minipage}[b]{0.5\linewidth}
\centering
\includegraphics[height=5.75in]{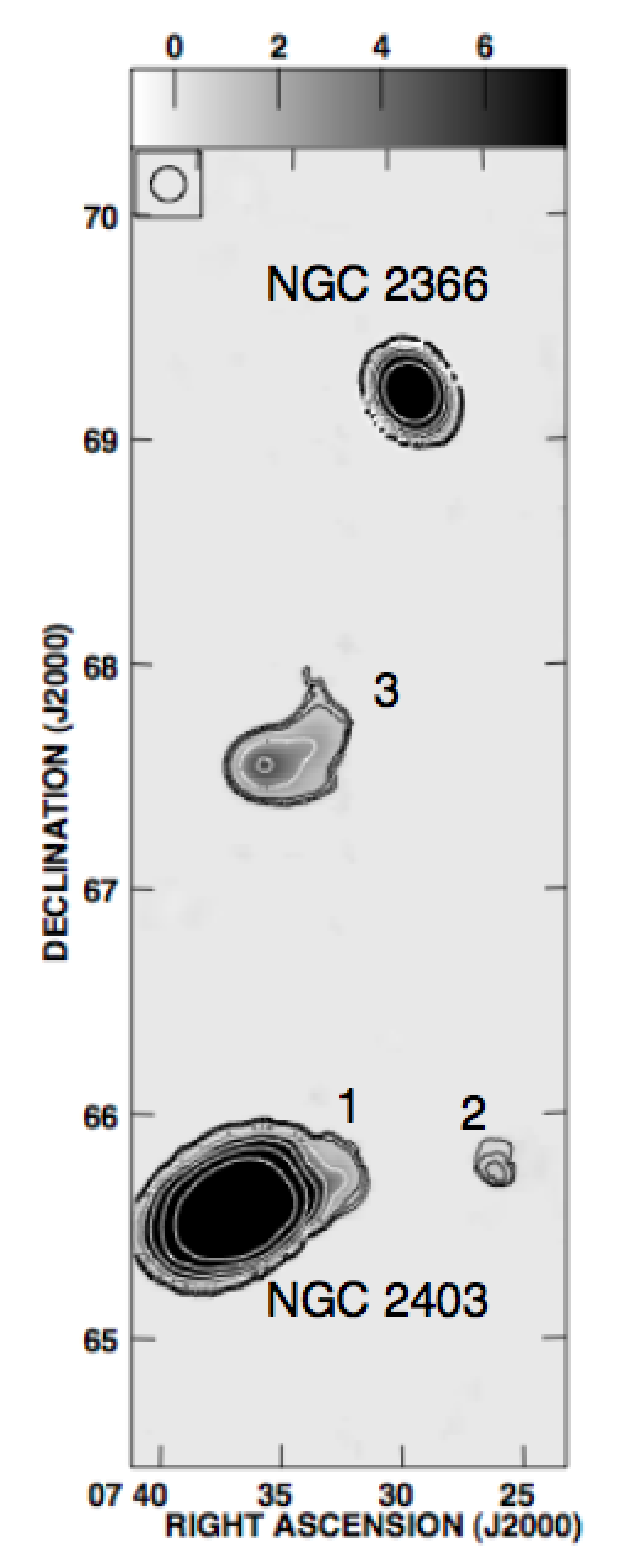}
\caption{NGC 2403 Group Region 1 integrated \hone column density map.  The 
galaxies are labeled and the newly discovered \hone feature are numbered. 
Velocity range: -260 to +335 km s$^{-1}$. Channels contaminated by
foreground gas are not included in the map. The greyscale is 
-0.75 to 7.5 Jy beam$^{-1}$ $\times$ km s$^{-1}$. Contours are at 5$\sigma$, 
7$\sigma$, 10$\sigma$, 25$\sigma$, 50$\sigma$, 100$\sigma$, 200$\sigma$.}
\label{mom01}
\end{minipage}
\hspace{2mm}
\begin{minipage}[b]{0.5\linewidth}
\centering
\begin{tabular}{c}
\includegraphics[height=2.75in]{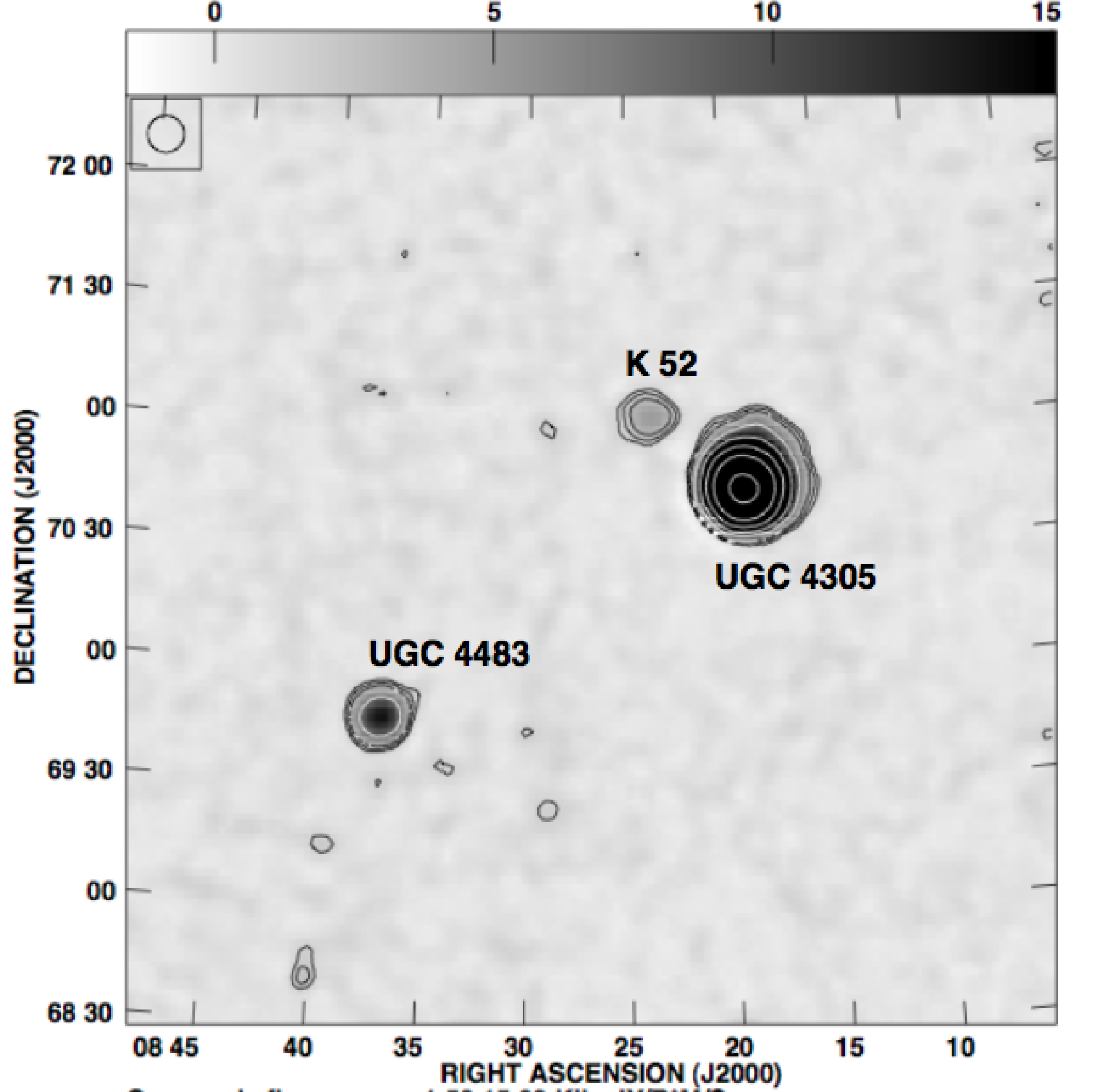} \\
\includegraphics[height=2.75in]{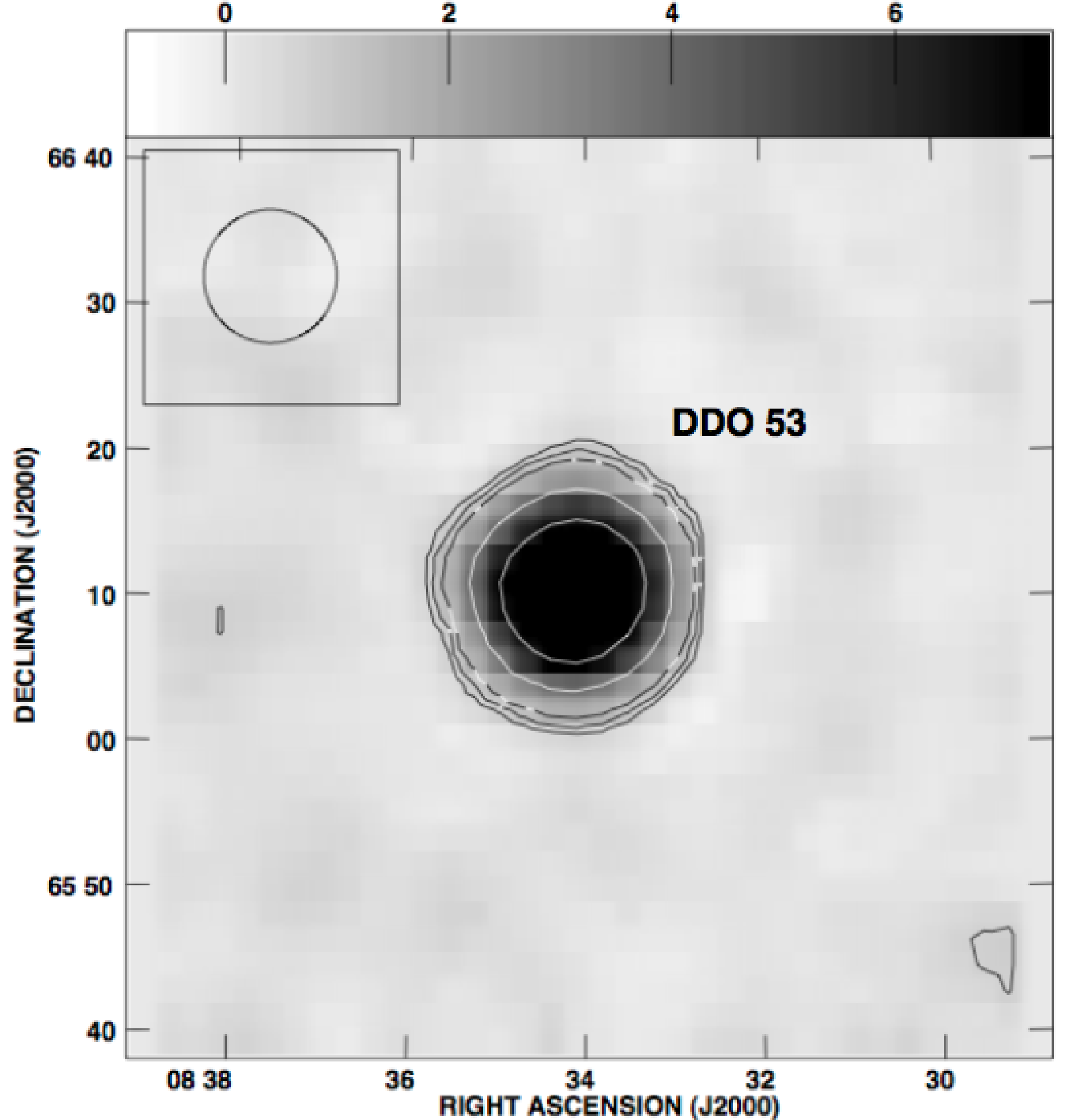}
\end{tabular}
\caption{Top: Region 2. Integrated \hone column density map,
80 to 220 km s$^{-1}$. Greyscale:  
-1.5 to 15 Jy beam$^{-1}$ $\times$ km s$^{-1}$. Contours are at 5$\sigma$, 
7$\sigma$, 10$\sigma$, 25$\sigma$, 50$\sigma$, 100$\sigma$, 200$\sigma$. Bottom: DDO
53. Integrated \hone column density map, 25 to 65 
km s$^{-1}$. Greyscale: -0.725 to 7.25 Jy beam$^{-1}$ $\times$ km s$^{-1}$. Contours are at 5$\sigma$, 
7$\sigma$, 10$\sigma$, 25$\sigma$, 50$\sigma$, 100$\sigma$, 200$\sigma$.}
\label{mom02}
\end{minipage}
\end{figure}

We placed their simulated components at the distance of NGC 2403 and at
the same relative location as is shown in Figure \ref{mom01}.   
After applying all observing constraints, we counted the number of objects
in the field.  
The simulation was performed with the Via Lactea II model oriented
in 4 different configurations  
The first configuration placed the Via Lactea II 
simulation components at Z=3300 kpc, the distance of NGC 2403, 
with velocity 131 km s$^{-1}$. The other three cases simply flipped the positions about the
principle axes.
In each configuration there were several dark matter halos that we should have detected in \hone.   An average of 8 components were
detected in the simulation configurations, with a range between 3 and
12 components.
The comparison of our observations and this simulation indicates that 
either the \hone mass fraction of the dark matter halos 
is less than 0.01 relative to total matter, 
or the Via Lactea II 
simulation is not an accurate model for the distribution of mass surrounding
the NGC 2403 group.

We compare the \hone characteristics of the M81/M82 galaxy group,
which is undergoing strong interactions between 4 large galaxies and has a
population of large \hone clouds and extended \hone, with the NGC 2403 galaxy 
group, which is not known to be interacting and contains no \hone clouds 
and no faint, extended \hone. 
Many galaxy groups have been searched for analogues to HVCs but none 
have been detected \citep{pis07}, except in galaxy groups with interacting
members \citep{kern07,chy08}. 

This paper is part of a larger project to observe \hone clouds in nearby galaxy groups. We plan 
 to obtain deeper observations of
the NGC 2403 group, and  are currently observing an expanded area around the M81 group and 3
more galaxy groups that have varying levels of ongoing galaxy interactions. We will compare this
dataset consisting of \hone cloud properties including mass, linewidth, velocity, and distance from
group galaxies to compare with the expected properties that result from
different \hone cloud formation mechanisms.

\acknowledgments

KMC wishes to thank the NRAO Graduate Summer Student Research Assistantship and
Pre-Doctoral Research programs for funding support. KMC also wishes to
acknowledge F.~Fraternali for helpful discussions regarding the comparison of
VLA and GBT data.

{\it Facilities:} \facility{GBT}.


\begin{thebibliography}{40}
\expandafter\ifx\csname natexlab\endcsname\relax\def\natexlab#1{#1}\fi
\expandafter\ifx\csname url\endcsname\relax
  \def\url#1{\texttt{#1}}\fi
\expandafter\ifx\csname urlprefix\endcsname\relax\def\urlprefix{URL }\fi
\providecommand{\eprint}[2][]{\url{#2}}

\bibitem[{{Begum} et~al.(2006){Begum}, {Chengalur}, {Karachentsev}, {Kaisin},
  \& {Sharina}}]{beg06}
{Begum}, A., {Chengalur}, J.~N., {Karachentsev}, I.~D., {Kaisin}, S.~S., \&
  {Sharina}, M.~E. 2006, \mnras, 365, 1220. \eprint{arXiv:astro-ph/0511253}

\bibitem[{{Ben Bekhti} et~al.(2006){Ben Bekhti}, {Br{\"u}ns}, {Kerp}, \&
  {Westmeier}}]{bekhti06}
{Ben Bekhti}, N., {Br{\"u}ns}, C., {Kerp}, J., \& {Westmeier}, T. 2006, \aap,
  457, 917. \eprint{arXiv:astro-ph/0607383}

\bibitem[{{Braun} \& {Burton}(1999)}]{braun99}
{Braun}, R., \& {Burton}, W.~B. 1999, \aap, 341, 437.
  \eprint{arXiv:astro-ph/9810433}

\bibitem[{{Braun} \& {Thilker}(2004)}]{braun04}
{Braun}, R., \& {Thilker}, D.~A. 2004, \aap, 417, 421.
  \eprint{arXiv:astro-ph/0312323}

\bibitem[{{Bregman}(1980)}]{bre80}
{Bregman}, J.~N. 1980, \apj, 236, 577

\bibitem[{{Bureau} \& {Carignan}(2002)}]{bureau02}
{Bureau}, M., \& {Carignan}, C. 2002, \aj, 123, 1316.
  \eprint{arXiv:astro-ph/0112325}

\bibitem[{{Chynoweth} et~al.(2008){Chynoweth}, {Langston}, {Yun}, {Lockman},
  {Rubin}, \& {Scoles}}]{chy08}
{Chynoweth}, K.~M., {Langston}, G.~I., {Yun}, M.~S., {Lockman}, F.~J., {Rubin},
  K.~H.~R., \& {Scoles}, S.~A. 2008, \aj, 135, 1983. \eprint{arXiv:0803.3631}

\bibitem[{{Diemand} et~al.(2008){Diemand}, {Kuhlen}, {Madau}, {Zemp}, {Moore},
  {Potter}, \& {Stadel}}]{die08}
{Diemand}, J., {Kuhlen}, M., {Madau}, P., {Zemp}, M., {Moore}, B., {Potter},
  D., \& {Stadel}, J. 2008, \nat, 454, 735. \eprint{0805.1244}

\bibitem[{{Fraternali} \& {Binney}(2006)}]{frat06}
{Fraternali}, F., \& {Binney}, J.~J. 2006, \mnras, 366, 449.
  \eprint{arXiv:astro-ph/0511334}

\bibitem[{{Fraternali} et~al.(2002){Fraternali}, {van Moorsel}, {Sancisi}, \&
  {Oosterloo}}]{frat02}
{Fraternali}, F., {van Moorsel}, G., {Sancisi}, R., \& {Oosterloo}, T. 2002,
  \aj, 123, 3124. \eprint{arXiv:astro-ph/0203405}

\bibitem[{{Grossi} et~al.(2008){Grossi}, {Giovanardi}, {Corbelli},
  {Giovanelli}, {Haynes}, {Martin}, {Saintonge}, \& {Dowell}}]{grossi08}
{Grossi}, M., {Giovanardi}, C., {Corbelli}, E., {Giovanelli}, R., {Haynes},
  M.~P., {Martin}, A.~M., {Saintonge}, A., \& {Dowell}, J.~D. 2008, \aap, 487,
  161. \eprint{0806.0412}

\bibitem[{{Kalberla} et~al.(2005){Kalberla}, {Burton}, {Hartmann}, {Arnal},
  {Bajaja}, {Morras}, \& {P{\"o}ppel}}]{kal05}
{Kalberla}, P.~M.~W., {Burton}, W.~B., {Hartmann}, D., {Arnal}, E.~M.,
  {Bajaja}, E., {Morras}, R., \& {P{\"o}ppel}, W.~G.~L. 2005, \aap, 440, 775.
  \eprint{arXiv:astro-ph/0504140}

\bibitem[{{Karachentsev} et~al.(2002){Karachentsev}, {Dolphin}, {Geisler},
  {Grebel}, {Guhathakurta}, {Hodge}, {Karachentseva}, {Sarajedini}, {Seitzer},
  \& {Sharina}}]{kar02}
{Karachentsev}, I.~D., {Dolphin}, A.~E., {Geisler}, D., {Grebel}, E.~K.,
  {Guhathakurta}, P., {Hodge}, P.~W., {Karachentseva}, V.~E., {Sarajedini}, A.,
  {Seitzer}, P., \& {Sharina}, M.~E. 2002, \aap, 383, 125

\bibitem[{{Karachentsev} \& {Kaisin}(2007)}]{kar07}
{Karachentsev}, I.~D., \& {Kaisin}, S.~S. 2007, \aj, 133, 1883.
  \eprint{arXiv:astro-ph/0701465}

\bibitem[{{Kere{\v s}} et~al.(2005){Kere{\v s}}, {Katz}, {Weinberg}, \&
  {Dav{\'e}}}]{keres05}
{Kere{\v s}}, D., {Katz}, N., {Weinberg}, D.~H., \& {Dav{\'e}}, R. 2005,
  \mnras, 363, 2. \eprint{arXiv:astro-ph/0407095}

\bibitem[{{Kern} et~al.(2007){Kern}, {Kilborn}, {Forbes}, \&
  {Koribalski}}]{kern07}
{Kern}, K., {Kilborn}, V., {Forbes}, D., \& {Koribalski}, B. 2007, ArXiv
  e-prints. \eprint{0711.1407}

\bibitem[{{Kravtsov} et~al.(2004){Kravtsov}, {Gnedin}, \& {Klypin}}]{kra04}
{Kravtsov}, A.~V., {Gnedin}, O.~Y., \& {Klypin}, A.~A. 2004, \apj, 609, 482.
  \eprint{arXiv:astro-ph/0401088}

\bibitem[{{Langston} \& {Turner}(2007)}]{lang07}
{Langston}, G., \& {Turner}, B. 2007, \apj, 658, 455

\bibitem[{{Larson} \& {Tinsley}(1978)}]{lar78}
{Larson}, R.~B., \& {Tinsley}, B.~M. 1978, \apj, 219, 46

\bibitem[{{Lockman} et~al.(2008){Lockman}, {Benjamin}, {Heroux}, \&
  {Langston}}]{loc08}
{Lockman}, F.~J., {Benjamin}, R.~A., {Heroux}, A.~J., \& {Langston}, G.~I.
  2008, \apjl, 679, L21. \eprint{0804.4155}

\bibitem[{{Lockman} et~al.(2002){Lockman}, {Murphy}, {Petty-Powell}, \&
  {Urick}}]{loc02}
{Lockman}, F.~J., {Murphy}, E.~M., {Petty-Powell}, S., \& {Urick}, V.~J. 2002,
  \apjs, 140, 331. \eprint{arXiv:astro-ph/0201039}

\bibitem[{{Maller} \& {Bullock}(2004)}]{mal04}
{Maller}, A.~H., \& {Bullock}, J.~S. 2004, \mnras, 355, 694.
  \eprint{arXiv:astro-ph/0406632}

\bibitem[{{Maloney} \& {Putman}(2003)}]{mal03}
{Maloney}, P.~R., \& {Putman}, M.~E. 2003, \apj, 589, 270.
  \eprint{arXiv:astro-ph/0302040}

\bibitem[{{Ott} et~al.(1994){Ott}, {Witzel}, {Quirrenbach}, {Krichbaum},
  {Standke}, {Schalinski}, \& {Hummel}}]{ott94}
{Ott}, M., {Witzel}, A., {Quirrenbach}, A., {Krichbaum}, T.~P., {Standke},
  K.~J., {Schalinski}, C.~J., \& {Hummel}, C.~A. 1994, \aap, 284, 331

\bibitem[{{Pisano} et~al.(2007){Pisano}, {Barnes}, {Gibson}, {Staveley-Smith},
  {Freeman}, \& {Kilborn}}]{pis07}
{Pisano}, D.~J., {Barnes}, D.~G., {Gibson}, B.~K., {Staveley-Smith}, L.,
  {Freeman}, K.~C., \& {Kilborn}, V.~A. 2007, \apj, 662, 959.
  \eprint{arXiv:astro-ph/0703279}

\bibitem[{{Putman} et~al.(2003){Putman}, {Staveley-Smith}, {Freeman}, {Gibson},
  \& {Barnes}}]{put03}
{Putman}, M.~E., {Staveley-Smith}, L., {Freeman}, K.~C., {Gibson}, B.~K., \&
  {Barnes}, D.~G. 2003, \apj, 586, 170. \eprint{arXiv:astro-ph/0209127}

\bibitem[{{Quilis} \& {Moore}(2001)}]{qui01}
{Quilis}, V., \& {Moore}, B. 2001, \apjl, 555, L95.
  \eprint{arXiv:astro-ph/0106253}

\bibitem[{{Sancisi} et~al.(2008){Sancisi}, {Fraternali}, {Oosterloo}, \& {van
  der Hulst}}]{san08}
{Sancisi}, R., {Fraternali}, F., {Oosterloo}, T., \& {van der Hulst}, T. 2008,
  \aapr, 15, 189. \eprint{0803.0109}

\bibitem[{{Simon} et~al.(2006){Simon}, {Blitz}, {Cole}, {Weinberg}, \&
  {Cohen}}]{simon06}
{Simon}, J.~D., {Blitz}, L., {Cole}, A.~A., {Weinberg}, M.~D., \& {Cohen}, M.
  2006, \apj, 640, 270. \eprint{arXiv:astro-ph/0511542}

\bibitem[{{Stil} et~al.(2006){Stil}, {Lockman}, {Taylor}, {Dickey}, {Kavars},
  {Martin}, {Rothwell}, {Boothroyd}, \& {McClure-Griffiths}}]{stil06}
{Stil}, J.~M., {Lockman}, F.~J., {Taylor}, A.~R., {Dickey}, J.~M., {Kavars},
  D.~W., {Martin}, P.~G., {Rothwell}, T.~A., {Boothroyd}, A.~I., \&
  {McClure-Griffiths}, N.~M. 2006, \apj, 637, 366.
  \eprint{arXiv:astro-ph/0509730}

\bibitem[{{Thilker} et~al.(2004){Thilker}, {Braun}, {Walterbos}, {Corbelli},
  {Lockman}, {Murphy}, \& {Maddalena}}]{thilker04}
{Thilker}, D.~A., {Braun}, R., {Walterbos}, R.~A.~M., {Corbelli}, E.,
  {Lockman}, F.~J., {Murphy}, E., \& {Maddalena}, R. 2004, \apjl, 601, L39.
  \eprint{arXiv:astro-ph/0311571}

\bibitem[{{Thom} et~al.(2006){Thom}, {Putman}, {Gibson}, {Christlieb}, {Flynn},
  {Beers}, {Wilhelm}, \& {Lee}}]{thom06}
{Thom}, C., {Putman}, M.~E., {Gibson}, B.~K., {Christlieb}, N., {Flynn}, C.,
  {Beers}, T.~C., {Wilhelm}, R., \& {Lee}, Y.~S. 2006, \apjl, 638, L97.
  \eprint{arXiv:astro-ph/0601139}

\bibitem[{{van der Hulst} \& {Sancisi}(1988)}]{vdH88}
{van der Hulst}, T., \& {Sancisi}, R. 1988, \aj, 95, 1354

\bibitem[{{van Woerden} et~al.(1999){van Woerden}, {Peletier}, {Schwarz},
  {Wakker}, \& {Kalberla}}]{vanw99a}
{van Woerden}, H., {Peletier}, R.~F., {Schwarz}, U.~J., {Wakker}, B.~P., \&
  {Kalberla}, P.~M.~W. 1999, in Stromlo Workshop on High-Velocity Clouds,
  edited by B.~K. {Gibson}, \& M.~E. {Putman}, vol. 166 of Astronomical Society
  of the Pacific Conference Series, 1

\bibitem[{{van Woerden} et~al.(2004){van Woerden}, {Wakker}, {Schwarz}, \& {de
  Boer}}]{vanW04}
{van Woerden}, H., {Wakker}, B.~P., {Schwarz}, U.~J., \& {de Boer}, K.~S.
  (eds.) 2004, {High Velocity Clouds}, vol. 312 of Astrophysics and Space
  Science Library

\bibitem[{{Wakker} \& {van Woerden}(1997)}]{wvw97}
{Wakker}, B.~P., \& {van Woerden}, H. 1997, \araa, 35, 217

\bibitem[{{Wakker} et~al.(2007){Wakker}, {York}, {Howk}, {Barentine},
  {Wilhelm}, {Peletier}, {van Woerden}, {Beers}, {Ivezi{\'c}}, {Richter}, \&
  {Schwarz}}]{wakker07}
{Wakker}, B.~P., {York}, D.~G., {Howk}, J.~C., {Barentine}, J.~C., {Wilhelm},
  R., {Peletier}, R.~F., {van Woerden}, H., {Beers}, T.~C., {Ivezi{\'c}}, {\v
  Z}., {Richter}, P., \& {Schwarz}, U.~J. 2007, \apjl, 670, L113.
  \eprint{arXiv:0710.3340}

\bibitem[{{Walter} et~al.(2002){Walter}, {Weiss}, \& {Scoville}}]{wal02}
{Walter}, F., {Weiss}, A., \& {Scoville}, N. 2002, \apjl, 580, L21.
  \eprint{arXiv:astro-ph/0210602}

\bibitem[{{Westmeier} et~al.(2008){Westmeier}, {Br{\"u}ns}, \& {Kerp}}]{west08}
{Westmeier}, T., {Br{\"u}ns}, C., \& {Kerp}, J. 2008, \mnras, 390, 1691.
  \eprint{0808.3611}

\bibitem[{{Yun} et~al.(1993){Yun}, {Ho}, \& {Lo}}]{yun93}
{Yun}, M.~S., {Ho}, P.~T.~P., \& {Lo}, K.~Y. 1993, \apjl, 411, L17

\end{thebibliography}

\begin{table}
\begin{center}
\caption{Known Galaxies in the NGC 2403 Group\tablenotemark{a}}
\begin{tabular}{l|c|c|c|c|c}
\tableline\tableline
Name 		& $\alpha$\tablenotemark{b}	& $\delta$\tablenotemark{b} & D\tablenotemark{c}	& V$_{hel}$\tablenotemark{c}	& Type	 \\ 
		& (J2000)		  & (J2000)		     & Mpc		& km s$^{-1}$		&	\\ \tableline
NGC 2403	& 07$^h$36$^m$52.7$^s$	& 65$^{\circ}$35$^`$52$^{``}$  & 3.30	& 131                 		& Sc           \\
DDO 44		& 07$^h$34$^m$11.9$^s$	& 66$^{\circ}$52$^`$58$^{``}$  & 3.19	& ...                     		& dSph    \\
NGC 2366 	& 07$^h$28$^m$55.7$^s$	& 69$^{\circ}$12$^`$59$^{``}$  & 3.19	& 99                 		& IBm             \\
UGC 4483 	& 08$^h$37$^m$03.1$^s$	& 69$^{\circ}$46$^`$44$^{``}$  & 3.21	& 156                		& BCD         \\
UGC 4305/HoII & 08$^h$19$^m$06.5$^s$	& 70$^{\circ}$43$^`$01$^{``}$  & 3.39	& 157                 		& Im             \\
KDG 52            	& 08$^h$23$^m$56.2$^s$	& 71$^{\circ}$01$^`$36$^{``}$  & 3.55 	& 113                 		& Irr       \\
DDO 53/UGC 4459& 08$^h$34$^m$08.6$^s$	& 66$^{\circ}$11$^`$03$^{``}$  & 3.56	& 20                 		& Irr            \\ 
VKN			& 08$^h$40$^m$11.1$^s$ 	& 68$^{\circ}$26$^`$11$^{``}$  & ...		& ...                     		& dSph(?)      \\
\tableline
\tablenotetext{a}{Within observed area}
\tablenotetext{b}{From \citet{kar02}}
\tablenotetext{c}{From \citet{kar07}}
\end{tabular}
\end{center}
\end{table}

\begin{table}
\begin{center}
\caption{Parameters of the Robert C. Byrd Green Bank Telescope System}
\begin{tabular}{ll}
\tableline\tableline
Telescope: \\
\hskip 3mm Diameter........................... & 100 m \\
\hskip 3mm Beamwidth (FWHM)................... &  9.1 \arcmin \\
\hskip 3mm Linear resolution ................. &  2.7 $D_{Mpc}$ kpc \\
Receiver: \\
\hskip 3mm System Temperature ................ & 20 K \\
Spectrometer: \\
\hskip 3mm Bandwidth.......................... & 12.5 MHz (2637.5 km s$^{-1}$) \\
\hskip 3mm Resolution, Hanning Smoothed....... & 6.1 kHz (1.29 km s$^{-1}$) \\
\tableline
\end{tabular}
\end{center}
\end{table}

\begin{table}
\begin{center}
\caption{Observations Summary}
\begin{tabular}{ll}
\tableline\tableline
Center Frequency (MHz)		& 1418.4065 \\
Bandwidth (MHz)			& 12.5 \\
Channel Width (kHz)			& 24.4 \\
Velocity Resolution (km s$^{-1}$)	& 5.2\\
Integration time (hours): 		& 21	\\
Typical RMS noise (mJy)		& 11\\
Sensitivity to \hone (7$\sigma$) & 2.2 $\times$ 10$^{6}$ M$_{\odot}$ \\
\end{tabular}
\end{center}
\end{table}

\begin{table}
\begin{center}
\caption{Calculated Galaxy Masses}
\begin{tabular}{l|c|c}
\tableline\tableline
Galaxy	& \hone Mass			& \hone Mass from \citet{kar07} \\
	& ($\times 10^9 M_{\odot}$)	&	 		\\ \tableline
NGC 2403		& 3.38 $\pm$ 0.26	& 3.31	 \\
DDO 44			& ...			& $<$ 0.001  \\
NGC 2366 		& 0.68 $\pm$ 0.05	& 0.71    \\
UGC 4483 		& 0.04 $\pm$ 0.005	& 0.03   \\
UGC 4305/HoII 		& 0.91 $\pm$ 0.07	& 0.98   \\
K 52            	& 0.02 $\pm$ 0.004 	& 0.01  \\
DDO 53/UGC 4459		& 0.05 $\pm$ 0.004	& ...    \\ 
VKN			& ...			& ...     \\
\tableline
\tableline
\end{tabular}
\end{center}
\end{table}

\begin{table}
\begin{center}
\caption{New \hone Cloud Properties}
\begin{tabular}{l|c|c|c|c|c|c}
\tableline\tableline
Cloud 		& $\alpha$	& $\delta$ & T$_{peak}$	& V$_{hel}$	&$\sigma_v$  & M$_{HI}$ ($\frac{D}{10 kpc})^{-2}$ \\ 
		& (J2000)      & (J2000)  & (K)	      & (km s$^{-1}$)		& (km s$^{-1}$) & (M$_{\odot}$)\\ \tableline
1	& 07$^h$32$^m$36.0$^s$ & 65$^{\circ}$43$^`$02$^{``}$ & 0.06& -150 & 20  & 14   \\
2	& 07$^h$25$^m$28.5$^s$ & 65$^{\circ}$49$^`$22$^{``}$ & 0.15& -114 & 20  & 57\\
3 	& 07$^h$36$^m$12.1$^s$ & 67$^{\circ}$33$^{``}$25$^{`}$ & 1.97& -205 & 30  &  475    \\
\tableline
\end{tabular}
\end{center}
\end{table}

\end{document}